\documentclass[11pt]{article}
\usepackage{jheppub}
\usepackage{ascmac}
\usepackage{amsthm}
\usepackage{physics}
\usepackage{appendix}
\usepackage{tikz}
\usepackage[edge length=0.7cm,root radius=1.2mm,mark=o]{dynkin-diagrams}
\usetikzlibrary{cd,backgrounds}

\DeclareMathOperator{\Ker}{Ker}
\newcommand{\W}{\mathcal{W}}
\newcommand{\g}{\mathfrak{g}}
\newcommand{\tp}{\widetilde{p}}
\newcommand{\te}{\widetilde{e}}

\arxivnumber{2609.14345}
\preprint{TIT/HEP-713}

\title{\boldmath Free field realization of
$\mathcal{W}$-algebra associated with exceptional Lie algebras}

\author{Daichi Ide,}
\author{Katsushi Ito}
\author{and Shigeki Miyazaki}
\affiliation{
Department of Physics, Institute of Science Tokyo,\\
Ookayama 2-12-1, Meguro-ku, Tokyo, 152-8551, Japan}

\emailAdd{ide.d.250c@m.isct.ac.jp}
\emailAdd{ito@phys.sci.isct.ac.jp}
\emailAdd{miyazaki.sgk@gmail.com}

\abstract{We study the free field realization of the $\mathcal{W}$-algebra associated with the exceptional Lie algebras $E_6$, $E_7$, $E_8$, and $F_4$. We develop a recursive construction in which a $\mathcal{W}$-algebra of rank $r$ is obtained from a $\mathcal{W}$-algebra of rank $r-1$ together with a free boson. The $\mathcal{W}$-currents are constructed from the zero commutation relation with the screening charges. The $\mathcal{W}E_6/\mathcal{W}E_7$ algebra is constructed from the $\mathcal{W}D_5/\mathcal{W}D_6$ algebra and is shown to be the same as that realized from the $\mathcal{W}A_5/\mathcal{W}E_6$ algebra, up to a change of the free field basis. The spin-$8$ generator of the $\mathcal{W}E_8$ algebra is built from the $\mathcal{W}D_7$ algebra. The recursive construction of the $\mathcal{W}BC_r$ algebras is also studied. We then realize the $\mathcal{W}F_4$ algebra based on the $\mathcal{W}BC_3$ algebra. Furthermore, the $\mathcal{W}$-charges of the generators of the $\mathcal{W}E_{6,7}$, $\mathcal{W}BC_{2,3}$, and $\mathcal{W}F_4$ algebras are calculated and expressed in terms of the Casimir invariants.}

\begin{document}
\maketitle

\section{Introduction}

The $\W$-algebra is an extension of the Virasoro algebra in two-dimensional conformal field theories by conserved currents with higher spin \cite{Zamolodchikov:1985wn,Fateev:1987vh}. The $\W$-algebras now play a central role in conformal field theory, integrable systems, representation theory, and supersymmetric gauge theories. One of the most systematic constructions of the $\W$-algebras is provided by the quantum Drinfeld–Sokolov reduction of the untwisted affine Lie algebra of a simple Lie algebra $\g$, denoted by $\W\g$ \cite{Bershadsky:1989mf,Feigin:1990pn,Feigin:1991wy,deBoer:1993iz}; see also \cite{Bouwknegt:1992wg} for a review. For classical Lie algebras of types $A$, $B$, $C$, and $D$, many properties of $\W$-algebras have been extensively investigated \cite{Fateev1988,Lukyanov:1989gg,Kausch:1991zm,Ito:1995ny}. The $\W$-algebras associated with $B_r$ and $C_r$ are equivalent and are denoted by $\W BC_r$. Explicit realizations of the $\W$-algebras in terms of the free fields are obtained by the Miura transformation and the quantum Drinfeld–Sokolov reduction, and their generators commute with the screening charges. Free field constructions have led to numerous applications, including the AGT correspondence \cite{Alday:2009aq,Wyllard:2009hg,Kanno:2013vi}, the ODE/IM correspondence \cite{Dorey:1998pt,Bazhanov:1998wj,Bazhanov:2001xm,Ito:2018eon,Ashok:2024zmw,Ito:2024kza,Ashok:2024ygp,Kudrna:2025bzg,Ide:2026cxp}, and exact results in supersymmetric quantum field theories \cite{Beem:2013sza,Cordova:2015nma,Xie:2019yds,Xie:2019vzr,Xie:2019zlb}.

By contrast, the $\W$-algebras associated with the exceptional Lie algebras $G_2$, $F_4$, $E_6$, $E_7$, and $E_8$ are considerably less known. The exceptional Lie algebras occupy a distinguished position in the classification of simple Lie algebras and possess remarkable algebraic and geometric structures that have no counterparts in the classical series. Their associated $\W$-algebras are therefore expected to exhibit equally rich structures, yet explicit constructions and computational tools remain comparatively limited. Among these cases, the $\W G_2$ and $\W E_6$ algebras have been constructed in previous works. For the $\W G_2$ algebra, the quantum Drinfeld-Sokolov reduction has been studied in \cite{Ito:1995ny}. The $\W E_6$ algebra was constructed based on the $\W A_5$ algebra \cite{Keller:2011ek}.

The motivation for the present work comes partly from the ODE/IM correspondence, in which the $\W$-algebras associated with simple Lie algebras govern the symmetries underlying quantum spectral problems \cite{Bazhanov:1994ft,Bazhanov:1996dr,Bazhanov:1996aq,Fioravanti:1995cq,Lukyanov:2006gv}. 
In particular, the coefficients of the WKB expansion of a certain period integral for the ODE correspond to the integrals of motion in conformal field theories with the $\W$-symmetries.
Although the cases of classical Lie algebras have been extensively studied \cite{Ito:2024kza}, the exceptional cases are a recent subject of study \cite{Ide:2026cxp}. The explicit constructions developed in this paper provide a computational foundation for studying the ODE/IM correspondence associated with exceptional Lie algebras \cite{Ito:2017ypt,Ito:2020htm}.

In this paper, we investigate the $\W$-algebras associated with exceptional Lie algebras of the $E_r$ and $F_4$ types from the perspectives of free field realizations and operator product expansions. Since the quantum Drinfeld-Sokolov reduction needs free fields whose number is the dimension of the Lie algebra, the computations of the $\W$-currents become quite cumbersome for the exceptional case. On the other hand, in the present construction, we need free fields whose number is equal to the rank of the Lie algebra. The generators of the $\W$-algebra in the free field realization are constructed by the zero commutation relation with the screening charges. We develop a method to explicitly compute $\W$-currents based on screening charges by decomposing the $\W$-algebra into another $\W$-algebra of one rank lower and a free boson. Our construction provides a practical framework for explicit calculations and is expected to be useful for applications to conformal field theory, integrable systems, and the ODE/IM correspondence.

The $\W A_r$ and $\W D_r$ algebras admit free field realizations based on quantum Miura transformations \cite{Fateev1988,Lukyanov:1989gg}. One of the important properties following from the Miura transformation is the recursive structure of the $\W$-currents. It is known that the $\W A_r$ algebra can be constructed recursively from the $\W A_{r-1}$ algebra plus a free boson \cite{Das:1991tq,Lu:1992yf}. This recursive structure is represented through $A_r\subset gl(r+1)$. The $\W D_r$ algebra can be constructed from the $\W D_{r-1}$ algebra and a free boson \cite{Lu:1992bn,Ito:2024kza}. We can represent these recursive structures by the following diagrams:

\begin{center}
    \begin{tikzcd}
        & \W gl(2) \ar[r]\ar[d] & \W gl(3) \ar[r] \ar[d]& \dots \ar[r] & \W gl(r+1) \ar[r] \ar[d] & \dots \\
        & \W A_1 & \W A_2 & \dots & \W A_r & \dots 
    \end{tikzcd}
\end{center}
and
\begin{center}
    \begin{tikzcd}
        & \W D_3=\W A_3 \ar[r] & \W D_4 \ar[r] & \dots \ar[r] & \W D_{r} \ar[r] & \dots .
    \end{tikzcd}
\end{center}
These sequences correspond to the recursive embeddings of the associated Dynkin diagrams\footnote{See Appendix~\ref{app:Lie}.} obtained by adding one node at each step.
Free field realizations based on deleting an arbitrary node of the Dynkin diagram were studied for $A_r$ and $D_r$ from the Miura transformation \cite{Lu:1992tr}.

This work aims to generalize this recursive structure of $\W$-algebras to the exceptional Lie algebras. We construct the $\W E_6$, $\W E_7$, and $\W E_8$ algebras starting from the $\W D_5$, $\W D_6$, and $\W D_7$ algebras, respectively, and adjoining a free boson. These constructions can be represented schematically by the following diagram:
\begin{center}
    \begin{tikzcd}
        \cdots\ar[r] & \W D_{4} \ar[r] &   \W D_5 \ar[r] \ar[rd] & \W D_6 \ar[r] \ar[rd] & \W D_7 \ar[r] \ar[rd] & \W D_8 \ar[r] &\cdots \\[-.3cm]
        & & & \W E_6 \ar[r] & \W E_7 \ar[r] & \W E_8 &
    \end{tikzcd}
\end{center}
which reflects the corresponding extensions of the Dynkin diagrams. In particular, the embedding structure of the Dynkin diagrams plays a crucial role in the construction of the $\W$-currents. These currents are defined as fields that commute with the screening charges associated with the simple roots. For the exceptional Lie algebra $E_{r+1}$, the $\W$-currents are constructed from those of $\W D_r$ and an additional free boson. The embedding of $D_r$ into $E_{r+1}$ implies that the simple roots can be decomposed into those of $D_r$ and their orthogonal direction. We will construct the full free field realization of the $\W$-currents for the $\W E_6$ algebra and present partial results for the $\W E_7$ and $\W E_8$ algebras.

We will find that the $\W BC_r$ algebras also have a recursive structure. The $\W G_2$ algebra, which has been constructed by the quantum Drinfeld–Sokolov reduction, can be understood as an extension of the $\W BC_1=\W A_1$ algebra. We also find that the $\W F_4$ algebra is constructed from the $\W BC_3$ algebra and a free boson. These structures are summarized as follows.
\begin{center}
    \begin{tikzcd}
        \W A_1=\W BC_1 \ar[r] \ar[rd] & \W BC_2\ar[r] & \W BC_3 \ar[r] \ar[rd] & \W BC_4 \ar[r] &\cdots \\[-.3cm]
        &\W G_2  & & \W F_4 & 
    \end{tikzcd}
\end{center}
we thus find a unified construction of the $\W$-algebras associated with the exceptional Lie algebras.

This paper is organized as follows: In Section~\ref{sec:WAandWD}, we briefly review the recursive construction of the $\W$-currents for the $\W A_r$ and $\W D_r$ algebras based on the quantum Miura transformations. In Section~\ref{sec:Wg}, we explain the general framework for the recursive construction of $\W$-algebras. In Section~\ref{sec:WE6}, we construct the free field realization of the $\W E_6$ algebra based on that of the $\W D_5$ algebra and show its equivalence to the result obtained from the $\W A_5$ algebra. In Section~\ref{sec:WE7}, we study the free field realization of the $\W E_7$ algebra based on the $\W D_6$ algebra. We also construct the spin-6 current from the $\W E_6$ algebra, which gives the same result as that obtained from the $\W D_6$ algebra. In Section~\ref{sec:WE8}, we obtain the spin-8 current of the $\W E_8$ algebra from the $\W D_7$ algebra. In Section~\ref{sec:WBC}, we study the recursive structure of the $\W BC_r$ algebras. In Section~\ref{sec:WF4}, we construct the $\W F_4$ algebra based on the $\W BC_3$ algebra. Section~\ref{sec:discussion} is devoted to the conclusion and discussion. The OPE computations are done with the help of the Mathematica program ``{\tt OPEdefs.m}" \cite{Thielemans:1991uw}.

\section{Recursive Construction of \texorpdfstring{$\mathcal{W}A_r$}{WAr} and \texorpdfstring{$\mathcal{W}D_r$}{WDr} Algebras}\label{sec:WAandWD}

In this section, we review some basic definitions of the $\W$-algebras associated with the simple Lie algebras $A_r$ and $D_r$, which have been extensively studied in the previous literature \cite{Fateev1988,Lukyanov:1989gg}.

Let $(z,\bar{z})$ be complex coordinates on the complex plane. A conformal field theory is defined by a set of local operators $\{\phi_i(z,\bar{z})\}$ and their operator product expansions (OPEs).
The set of operators is organized into the representation of the conformal algebra that is generated by the energy-momentum tensor $T(z)$ and its extension obtained by adding some extra generators. The $\W$-algebra is an example of such an extended conformal algebra by introducing additional higher-spin operators.

The energy-momentum tensor $T(z)$ satisfies the OPE
\begin{equation}
    T(z)T(w)
    = \frac{c/2}{(z-w)^4} + \frac{2T(w)}{(z-w)^2} + {\frac{\partial T(w)}{z-w}} + \cdots,
\end{equation}
where $c$ is the central charge. The term $\cdots$ denotes the regular terms. The primary field $\phi(z)$ with conformal dimension $\Delta$ has the OPE of the form:
\begin{equation}
    T(z)\phi(w)
    = \frac{\Delta \phi(w)}{(z-w)^2} + \frac{\partial \phi(w)}{z-w} + \cdots.
\end{equation}
The OPEs between $\W$-currents in the $\W$-algebra contain non-linear terms involving normal-ordered products of the currents. For two operators $A(z)$ and $B(z)$ of conformal dimensions $\Delta_A$ and $\Delta_B$, respectively, we write their OPE as
\begin{equation}\label{eq:OPEofAB}
    A(z)B(w)
    = \sum_{1\leq k\leq \Delta_A+\Delta_B} \frac{\{AB\}_k(w)}{(z-w)^k} + \cdots.
\end{equation}
We also define the normal-ordered product of $A(z)$ and $B(z)$ by
\begin{equation}
    (AB)(w)
    = \frac{1}{2\pi i} \oint_{C_w}\dd{z} \frac{A(z)B(w)}{z-w},
\end{equation}
where the contour $C_w$ surrounds the point $w$ in the complex plane.

We next introduce free scalar fields $\phi_i(z)$ ($i=1,2,\dots$), satisfying
\begin{equation}
    \phi_i(z)\phi_j(w)
    = -\delta_{ij}\log(z-w) + \cdots.
\end{equation}
It is also convenient to introduce $p_i(z)=i\partial\phi_i(z)$, which satisfy
\begin{equation}
    p_i(z)p_j(w)
    = \frac{\delta_{ij}}{(z-w)^2} + \cdots.
\end{equation}
Based on these notations and conventions, we first introduce the $\W A_r$ algebra.

\subsection{\texorpdfstring{$\mathcal{W}A_r$}{WAr} Algebra}

The $\W A_r$ algebra is generated by currents $W_2,W_3,\ldots, W_{r+1}$, where $W_s$ denotes a current of spin $s$. $W_2$ is identified with the energy-momentum tensor. Its free field realization is defined by the quantum Miura transformation involving $r$ scalar fields $\phi=(\phi_1,\dots,\phi_r)$ \cite{Fateev1988}:
\begin{equation}
    (Q\partial)^{r+1} - \sum_{k=2}^{r+1}W_k(z) (Q \partial)^{r-k+1}
    = ~:(Q \partial-h_1\cdot i \partial \phi(z)) \cdots (Q \partial-h_{r+1} \cdot i \partial \phi(z)):~,
\end{equation}
where $h_i$ are the weight vectors of the fundamental representation of $A_r$. The symbol $:\ :$ denotes normal-ordered product. We have
\begin{equation}
    T(z) = W_2(z)
    = -\frac{1}{2}(\partial\phi \partial\phi)(z) - i Q\rho \cdot \partial^2 \phi(z)
\end{equation}
where $\rho = \sum_{i=1}^{r+1}(r+1-i)h_i$ is the Weyl vector of $A_r$. The central charge is $c(A_r)=r-12Q^2\rho^2$, where $\rho^2 = \frac{1}{12}r(r+1)(r+2)$. We can construct the higher-spin $\W$-generators recursively. First, we begin with the quantum Miura transformation associated with $gl(r)$:
\begin{equation}
    {\cal L}^r
    = ~:(Q\partial -\tp_r)\dots (Q\partial -\tp_1):~=(Q\partial)^r -\sum_{k=1}^{r}f^{(r)}_k (Q\partial)^{r-k},
\end{equation}
where $\tp_i = i\partial \widetilde{\phi}_i$ for free bosons $\widetilde{\phi}_i$ with the OPE $\widetilde{\phi}_i(z)\widetilde{\phi}_j(w) = -\delta_{ij}\log(z-w)+\cdots$. It is related to the operator for $gl(r+1)$ by ${\cal L}^{r+1} =~:(Q\partial -\tp_{r+1}){\cal L}^r:$. This leads to the relations
\begin{equation}
    f^{(r+1)}_k
    = f_{k}^{(r)} + Q\partial f^{(r)}_{k-1} - (\tp_{r+1} f^{(r)}_{k-1}), \quad (k=1,\dots, r+1),
\end{equation}
where $f^{(r)}_0=-1$ and $f^{(r)}_{r+1}=0$. 
If we substitute $\tp_i\to h_{r+2-i}\cdot p$ so that $\sum_{i=1}^{r+1}\tp_i=0$, we obtain the $\W$-generators $W_s$ as $f^{(r+1)}_s$ for $A_r$.

The $\W$-currents are shown to commute with the screening charges $Q_i^{\pm} = \oint dz S_{\alpha_i}^{\pm}(z)$, where
\begin{equation}\label{eq:screening1}
    S_{\alpha_i}^{\pm}(z)
    = ~: e^{-i \alpha_{\pm}\alpha_i\cdot \phi} :, \quad (i=1,\dots,r),
\end{equation}
and $\alpha_i$ are the simple roots, $\alpha_{+}=b$, $\alpha_{-}= -\frac{1}{b}$ with $Q=b-\frac{1}{b}$.

\subsection{\texorpdfstring{$\mathcal{W}D_r$}{WDr} Algebra}

We next consider the $\W D_r$ algebra. Let $e_1,\dots,e_r$ be an orthonormal basis of ${\mathbb R}^r$. The simple roots of the Lie algebra of type $D_r$ are
\begin{equation}
    \alpha_i = e_i - e_{i+1}, \;\; (i=1, \dots, r-1), \qquad
    \alpha_r = e_{r-1} + e_r.
\end{equation}
The fundamental weights are given by
\begin{align}
    \Lambda_i &= e_1 + \cdots + e_i, \quad (i=1,\dots, r-2), \nonumber\\
    \Lambda_{r-1} &= \frac{1}{2}(e_1+\cdots+e_{r-1}-e_r), \quad
    \Lambda_r = \frac{1}{2}(e_1+\cdots+e_{r-1}+e_r).
\end{align}
We introduce $r$ free bosons $\phi=(\phi_1,\dots,\phi_r)$ and $p=\mathrm{i}\partial \phi$. The $\W$-algebra is generated by currents of spins $2,4,\dots,2r-2$, denoted by $W_2,W_4,\dots,W_{2r-2}$, together with a spin-$r$ current $R_r$. The free field realization of $R_r$ is given by
\begin{equation}\label{eq:wd_gen1}
    R_r(z)
    = ~: (Q \partial - p_1(z)) \cdots (Q\partial - p_r(z)):\cdot~1.
\end{equation}

The remaining generators are defined through the OPE of $R_r$ with itself:
\begin{equation}\label{eq:OPEofRR}
    R_r(z)R_r(w)
    = (z-w)^{-2r}A_r + \sum_{k=1}^{r-1}(z-w)^{-2(r-k)}A_{r-k} (W_{2k}(z) + W_{2k}(w)) + \cdots,
\end{equation}
where 
\begin{equation}
    A_k
    = \prod_{j=1}^{k-1}(1-2j(2j+1)Q^2).
\end{equation}
The energy-momentum tensor $T(z)=W_2(z)$ is given by
\begin{equation}\label{eq:WAemtensor}
    T(z)
    = -\frac{1}{2}(\partial\phi\cdot\partial\phi)(z) - i Q\rho\cdot \partial^2 \phi,
\end{equation}
where $\rho=\sum_{i=1}^{r}(r-i)e_i$ is the Weyl vector. The central charge is $c(D_r)=r-r(2r-1)(2r-2)Q^2$.

In \cite{Lu:1992bn,Ito:2024kza}, the recursive structure of the $\W$-algebra has been studied based on \eqref{eq:wd_gen1}, where the spin-$r$ current $R_r^{(r)}$ of the $\W D_r$ algebra and the spin-$(r-1)$ current $R^{(r-1)}_{r-1}$ of the $\W D_{r-1}$ algebra obey the relation:
\begin{equation}\label{eq:frrec1}
    R^{(r)}_{r}
    = -(\tp_{r} R^{(r-1)}_{r-1}) + Q \partial R^{(r-1)}_{r-1}.
\end{equation}
Here we have defined $\tp_{i}=p_{r+1-i}$ and $\tp_1, \dots, \tp_{r-1}$ are the free fields for the $\W D_{r-1}$ algebra. 
Based on this relation and the OPE \eqref{eq:OPEofRR}, one can derive the recurrence relations for the other $\W$-currents. Their explicit forms are presented in Appendix~\ref{app:WD-recursion}. The $\W$-currents commute with the screening charges associated with \eqref{eq:screening1}. The free field realization of the $\W D_r$ algebra developed here provides a useful starting point for constructing the $\W$-currents of the $\W E_{r+1}$ algebra.

\section{Recursive Construction of \texorpdfstring{$\mathcal{W}\mathfrak{g}$}{Wg} Algebra}\label{sec:Wg}

In Section~\ref{sec:WAandWD}, we reviewed the recursive construction of the $\W A_r$ and $\W D_r$ algebras, in which a $\W$-algebra of rank $r$ is constructed from a $\W$-algebra of rank $r-1$ together with an additional free boson. Motivated by these examples, we now describe a general framework for the recursive construction of $\W$-currents based on screening charges.

Let $\g_r$ be a simple Lie algebra of rank $r$ with simple roots $\alpha_i(\g_r)\;(i=1,\ldots,r)$. We normalize the invariant bilinear form so that the long roots have squared length $2$, and we define the coroot associated with a root $\alpha$ by $\alpha^\vee = 2\alpha/(\alpha,\alpha)$. $\Lambda_i(\g_r)$ denote the fundamental weights which satisfy $\Lambda_i(\g_r)\cdot \alpha_j^{\vee}(\g_r)=\delta_{ij}$.

The $\W\g_r$ algebra is an extension of the Virasoro algebra generated by the energy-momentum tensor $W_2$ and higher-spin currents $W_s$, where the spins $s$ are given by the degrees of the independent Casimir invariants of $\g_r$. In the free field realization based on $r$ free bosons $\phi_i$ with $p_i = i\partial{\phi_i}$, the energy-momentum tensor is obtained from the quantum Drinfeld–Sokolov reduction as
\begin{equation}\label{eq:EMtensorWg}
    W_2(z) = 
    -\frac{1}{2}(\partial{\phi(z)}\cdot\partial{\phi(z)})
    - \qty( b\rho(\g_r) - \frac{1}{b}\rho^{\vee}(\g_r) ) \cdot i\partial^{2}{\phi(z)}.
\end{equation}
Here, $\rho(\g_r)$ and $\rho^\vee(\g_r)$ are the Weyl vector and the co-Weyl vector of $\g_r$, defined by the sum of the fundamental (co-)weights, respectively, and $b$ is assumed to be generic. The corresponding central charge is
\begin{equation}
    c(\g_r) = 
    r - 12\qty( b\rho(\g_r) - \frac{1}{b}\rho^{\vee}(\g_r) )^{2}.
\end{equation}
For simply-laced $\g_r$, we have $\rho^\vee = \rho$, and we introduce $Q = b - b^{-1}$. 

The higher-spin currents are characterized by the OPEs with the screening currents. For each simple root $\alpha_i(\g_r)$, we introduce the screening currents:
\begin{equation}\label{eq:scr}
    S_{\alpha_i}^{+}(z) = ~:e^{-ib\alpha_{i}(\g_r)\cdot\phi(z)}:~,
    \quad
    S_{\alpha_i}^{-}(z) = ~:e^{i\frac{1}{b}\alpha_{i}^{\vee}(\g_r)\cdot\phi(z)}:.
\end{equation}
These screening currents have conformal weight one with respect to $W_2$. We also introduce the corresponding screening charges:
\begin{equation}\label{eq:scrQ}
    Q_i^{\pm} = \oint \dd{z} S_{\alpha_i}^{\pm}(z).
\end{equation}
In the free field realization, the $\W\g_r$ algebra is identified with the common kernel of the screening charges \cite{Feigin:1991wy,Genra_2017}:
\begin{equation}\label{eq:commonkernel}
    \W\g_r = \bigcap_{i=1}^{r} \qty(\Ker{Q_i^{+}} \cap \Ker{Q_i^{-}}),
\end{equation}
where $\Ker{Q_i^{\pm}}$ denotes the space of currents commuting with $Q_i^{\pm}$. Equivalently, every generator $W_s(w)$ of the $\W\g_r$ algebra satisfies the zero commutation relation with $Q_i^{\pm}$:
\begin{equation}
\label{eq:zerocom}
    [Q_i^{\pm},W_s(w)] = 0 \qquad (i=1,\ldots,r).
\end{equation}
From the definition of the screening charges in \eqref{eq:scrQ} and the notation for the OPE coefficients introduced in \eqref{eq:OPEofAB}, this is equivalent to
\begin{equation}\label{eq:condition}
    \{S_{\alpha_{i}}^{\pm} \, W_s\}_1 = 0 \qquad (i=1,\ldots,r),
\end{equation}
which are the zero commutation relations associated with the simple roots $\alpha_i$ of $\g_r$. Since $S_{\alpha_i}^{\pm}$ is a primary field of conformal weight one, $W_2$ satisfies \eqref{eq:condition} as expected.
In \cite{Watts:1992he}, the isomorphism between the Hilbert space of the $\W\g_r$ algebra and that of the $\W$-algebra with lower rank and free bosons was studied.

We now construct the generators of the $\W\g_r$ algebra explicitly, which belong to the kernel of \eqref{eq:commonkernel}, by expressing them in terms of those of the $\W\g_{r-1}$ algebra. Among the simple roots of $\g_r$, we choose one, denoted by $\alpha_\ast$, such that the remaining $r-1$ simple roots form the simple root system of a simple subalgebra $\g_{r-1}$ of rank $r-1$. Its Dynkin diagram is obtained by removing the node corresponding to $\alpha_\ast$ from that of $\g_r$. Let $w_s$ denote the generators of the $\W\g_{r-1}$ algebra, which satisfy the zero commutation relations \eqref{eq:zerocom} associated with the simple roots of $\g_{r-1}$. We then introduce an additional free boson $\varphi$ and define $p = i\partial{\varphi}$, where the vector corresponding to $\varphi$ in the root space of $\g_{r}$ is orthogonal to the root subspace of $\g_{r-1}$. We normalize $\varphi$ as follows
\begin{equation}
    \varphi(z)\varphi(w) = -\log{(z-w)} + \cdots.
\end{equation}
Consequently, the OPE of $p$ with the screening currents associated with $\g_{r-1}$ has no singularity. To construct a spin-$s$ current $W_s$ with $s>2$ of the $\W\g_{r}$ algebra, we take the most general ansatz:
\begin{equation}
    W_s(z) = \sum_j x_j X_j(p(z),\{w_k(z)\}),
\end{equation}
where $X_j$ are spin-$s$ composite fields built from $p, w_k$ and their derivatives, and $x_j$ are the coefficients to be determined. Since $p$ and $w_k$ satisfy the zero commutation relations associated with the simple roots of $\g_{r-1}$, the above $W_s$ automatically satisfies these relations. Therefore, it remains only to impose the zero commutation relation associated with the additional simple root $\alpha_{\ast}$:
\begin{equation}
    \{S_{\alpha_\ast}^{\pm} \, W_s\}_1 = 0.
\end{equation}
For simply-laced $\g_r$, the transformation $b \to - b^{-1}$ exchanges $S_{\alpha_{\ast}}^+$ and $S_{\alpha_{\ast}}^-$ while leaving $Q=b - b^{-1}$ invariant. Since the coefficients of the $\W$-currents are expressed in terms of $Q$, the zero commutation condition associated with $S_{\alpha_{\ast}}^-$ is obtained from that for $S_{\alpha_{\ast}}^+$ by this transformation. It is sufficient to impose the condition for $S_{\alpha_{\ast}}^+$. For the non-simply laced $\g_r$, the two screening currents are not related by this transformation in the same way, and both conditions must be imposed independently.

This relation yields a set of linear equations satisfied by the coefficients $x_j$. Solving these equations constrains the coefficients but in general does not determine them completely. As we shall see later, the undetermined coefficients correspond to a general normalization of $W_s$ and the coefficients of the terms built from the lower-spin generators.

Let us apply the above method to the construction of the $\W D_r$ algebra from the $\W D_{r-1}$ algebra as an example. We decompose the simple roots of $D_r$ in terms of those of $D_{r-1}$ and its orthogonal direction denoted by $e_{1}$ as 
\begin{equation}\label{eq:Drdecom}
    \alpha_1(D_r) = -\Lambda_1(D_{r-1}) + e_1,
    \quad
    \alpha_i(D_r) = \alpha_{i-1}(D_{r-1}) ~~ (i=2,\ldots,r),
\end{equation}
where $\Lambda_i(D_{r-1})$ are the fundamental weights of $D_{r-1}$. Under this decomposition $\alpha_\ast$ is $\alpha_1(D_r)$. The additional free boson $p$ lies in the direction orthogonal to the root space of $D_{r-1}$. If we identify it with $\tp_r$, the recursion relation in \eqref{eq:frrec1} agrees with the one following from the zero commutation relation associated with $\alpha_\ast$. The spin-2 current $W_2^{(r)}$ is decomposed as in Eq.~\eqref{eq:wdrw2decomposition} in Appendix~\ref{app:WD-recursion}. The most general spin-$4$ ansatz consists of eleven terms written built from $W^{(r-1)}_2$, $W^{(r-1)}_4$ obtained from the Miura transformation, $\tp_r$, and their derivatives. Imposing the zero commutation relation associated with $\alpha_\ast$ leaves three coefficients undetermined, and we obtain
\begin{equation}
    W^{(r)}_4 = x_1\widehat{W}^{(r)}_4 + x_2 (W^{(r)}_2 W^{(r)}_2) + x_3 \partial^2 W^{(r)}_2
\end{equation}
where
{\small
\begin{align}
    \widehat{W}^{(r)}_4
    &= W^{(r-1)}_4 + (\tp_r(\tp_r W^{(r-1)}_2)) - 2(r-2)Q(\partial\tp_r W^{(r-1)}_2) - Q(\tp_r \partial W^{(r-1)}_2) - (r-2)Q^2(\tp_r \partial^2\tp_r)
    \nonumber\\
    &\quad - \frac{1}{4}(1-Q^2(2r-4)(2r-5))(\partial\tp_r \partial\tp_r) + \frac{Q}{12}(r-1)(1-Q^2(2r-4)(2r-9))\partial^3\tp_r.
\end{align}
}
For $(x_1,x_2,x_3)=(1,0,(r-2)Q^2)$, this result agrees with the recursion relation obtained from the Miura transformation.

We can also construct the $\W A_r$ algebra directly from the $\W A_{r-1}$ algebra. 
We decompose the simple roots of $A_r$ in terms of those of $A_{r-1}$ and its orthogonal direction denoted by $e_r$ as
$\alpha_i(A_r)=\alpha_i(A_{r-1})~(i=1,\dots,r-1$) and $\alpha_{r}(A_r)=-\Lambda_{r-1}(A_{r-1}) +\sqrt{\frac{r+1}{r}}e_r$, where $\Lambda_i(A_{r-1})$ are the fundamental weights of $A_{r-1}$. We take $\alpha_{\ast}$ as $\alpha_{r}(A_r)$.  
We also introduce an additional free boson $p_r$ along the $e_r$ direction.

The spin-2 current $W_2^{(r)}$ obeys the recurrence relation
\begin{equation}
    W_2^{(r)} = W_2^{(r-1)}+\frac{1}{2}(p_r p_r)-Q\frac{\sqrt{r(r+1)}}{2}\partial p_r,
\end{equation}
which reproduces Eq.~\eqref{eq:WAemtensor}. The spin-3 current $W_3^{(r)}$ is obtained by imposing the zero commutation relation, which is in agreement with the result of the Miura transformation. For $r=2$, the current $W_3^{(2)}$ corresponds to the spin-$3$ current of \cite{Fateev:1987vh}.

Finally, we comment on the $\W G_2$ algebra. It is obtained from the decomposition of the simple roots of $G_2$ in terms of that of $A_1=BC_1$ and its orthogonal direction specified by the unit vector $e_2$:
$\alpha_1(G_2)=\alpha_1(A_1)$ and $\alpha_2(G_2)=-\Lambda_1(A_1)+\frac{1}{\sqrt{6}}e_2$.  The spin-2 and spin-6 currents can be constructed from the zero commutation relations with the screening charges, and agree with those obtained from the quantum Drinfeld-Sokolov reduction \cite{Ito:1995ny}.

\section{Free Field Realization of \texorpdfstring{$\mathcal{W}E_6$}{WE6} Algebra}\label{sec:WE6}

The $\W E_6$ algebra is generated by the $\W$-currents $W_s$ of spins $s=2,5,6,8,9$, and $12$. A free field realization of the $\W E_6$ algebra was discussed in \cite{Keller:2011ek}, where the $\W$-currents of the $\W E_6$ algebra are expressed in terms of the $\W$-currents of the $\W A_5$ algebra and a free boson. The $\mathbb{Z}_2$ outer-automorphism structure of $E_6$ implies that the currents can be chosen to have definite parity under the $\mathbb{Z}_2$-automorphism. This property considerably simplifies the formula for the spin-5 current. This simplification is not available for $E_7$ and $E_8$, whose outer-automorphism groups are trivial. In this work, we instead take the $D_5$ subalgebra as the starting point and construct the $\W E_6$ algebra from the $\W D_5$ algebra. The two constructions of $\W E_6$ are related by a change of the free field basis.

\subsection{Construction from \texorpdfstring{$\mathcal{W}D_5$}{WD5} Algebra}

We now construct the generators of the $\W E_6$ algebra from those of the $\W D_5$ algebra and a free boson. The Dynkin diagrams of $D_5$ and $E_6$ are found in Appendix~\ref{app:Lie}. In terms of the simple roots of $D_5$, those of $E_6$ are expressed as
\begin{equation}\label{eq:E6root}
    \alpha_i(E_6) = \alpha_i(D_5), ~~ (i=1,\ldots,4), \quad
    \alpha_5(E_6) = -\Lambda_4(D_5) + \frac{\sqrt{3}}{2}e_6, \quad
    \alpha_6(E_6) = \alpha_5(D_5),
\end{equation}
where $e_6$ is a unit vector orthogonal to the root space of $D_5$, and $\alpha_5$ is identified with the additional simple root $\alpha_\ast$ introduced in Section~\ref{sec:Wg}. For the above choice of the simple roots, the Weyl vector of $E_6$ takes the form:
\begin{equation}\label{eq:E6weyl}
\rho(E_6)=\rho(D_5)+4\sqrt{3}e_6.
\end{equation}
Under this embedding of the $D_5$ root system into $E_6$, the screening currents $S_{\alpha_i}^{\pm}~(i \neq 5)$ are identified with those of the $\W D_5$ algebra. The generators $w_s~(s=2,4,6,8)$ and $r_5$ of the $\W D_5$ algebra already satisfy the zero commutation relations for $Q^{\pm}_i~(i \neq 5)$.

Let $\phi_i~(i=1,\ldots,5)$ denote the free bosons in the free field realization of the $\W D_5$ algebra. Following Section~\ref{sec:Wg}, we introduce an additional free boson $\varphi$ in the $e_6$ direction and define $p=i\partial{\varphi}$. Since $\alpha_i~(i\neq5)$ lie in the root space of $D_5$, $p$ satisfies the zero commutation relations for $Q^{\pm}_i~(i\neq5)$. Therefore, any $\W$-current $W_s$ of the $\W E_6$ algebra constructed from $r_5, w_s$, and $p$ satisfies the zero commutation relations for $Q^{\pm}_i~(i \neq 5)$. We then impose the remaining zero commutation relation for $Q^{\pm}_5$:
\begin{equation}\label{eq:screening cond. of E6}
    \{S_{\alpha_5}^{\pm} \, W_s\}_1 = 0.
\end{equation}
Since $E_6$ is simply laced, the relations for $S_{\alpha_5}^+$ and $S_{\alpha_5}^-$ are equivalent as explained in Section~\ref{sec:Wg}, so only the relation for $S_{\alpha_5}^{+}$ has to be imposed.

Let us begin with the spin-$2$ current $W_2$ of the $\W E_6$ algebra. It can be determined directly from \eqref{eq:EMtensorWg} together with the decomposition \eqref{eq:E6weyl}:
\begin{equation}\label{eq:EMtensorWE6}
    W_2 = w_2 + \frac{1}{2}(pp) - 4\sqrt{3}Q\partial{p},
\end{equation}
which has the central charge $c(E_6) = 6 - 936Q^2$.

Before considering the spin-5 current, we examine the most general ansatz for the currents of spins 3 and 4. Imposing the zero commutation relation \eqref{eq:screening cond. of E6}, we find that the solutions contain only fields constructed from the lower-spin current $W_2$ and its derivatives. Thus, no independent generators of spins 3 and 4 arise, as expected from the spin content of the $\W E_6$ algebra.

We next consider the spin-$5$ current $W_5$ of the $\W E_6$ algebra. The most general ansatz for $W_5$ can be written as
\begin{equation}
\begin{aligned}
    W_5 &= x_1 r_5 + x_2 (p w_4) + x_3 \partial{w_4} + x_4 (p(w_2 w_2)) + x_5 (p(p(p w_2))) + x_6 (p(\partial{p} w_2)) + x_7 (\partial^{2}{p} w_2) \\
    &\quad + x_8 (w_2 \partial{w_2}) + x_9 (p(p \partial{w_2})) + x_{10} (\partial{p} \partial{w_2}) + x_{11} (p \partial^{2}{w_2}) + x_{12} \partial^{3}{w_2} + x_{13} (p(p(p(p p)))) \\
    &\quad + x_{14} (p(p(p \partial{p}))) + x_{15} (p(\partial{p} \partial{p})) + x_{16} (p(p \partial^{2}{p})) + x_{17} (\partial{p} \partial^{2}{p}) + x_{18} (p \partial^{3}{p}) + x_{19} \partial^{4}{p}.
\end{aligned}
\end{equation}
Imposing the zero commutation relation \eqref{eq:screening cond. of E6} yields a set of linear equations obeyed by the coefficients $x_j~(j=1,\ldots,19)$. Solving these equations determines all the coefficients except $x_1$, $x_8$, and $x_{12}$ as
\begin{equation}
\begin{aligned}
    x_2 &= \frac{x_1}{\sqrt{3}}, \quad x_3 = -Qx_1, \quad x_4 = -\frac{x_1}{2 \sqrt{3}}, \quad x_5 = -\frac{x_1}{6 \sqrt{3}}, \quad x_6 = x_8, \\
    x_7 &= \frac{2 Q \left(5Q x_1 - 6 x_8\right)}{\sqrt{3}}, \quad x_9 = \frac{x_8}{2}, \quad x_{10} = \sqrt{3} Q \left(3Q x_1 - 4 x_8\right), \quad x_{11} = \frac{\left(1-24 Q^2\right)x_1}{4 \sqrt{3}}, \\
    x_{13} &= \frac{x_1}{24 \sqrt{3}}, \quad x_{14} = \frac{1}{6} \left(3 x_8 - 8Q x_1\right), \quad x_{15} = \frac{\sqrt{3} Q\left(13Q x_1 - 8 x_8\right)}{2}, \\
    x_{16} &= \frac{Q \left(11Q x_1 - 6 x_8\right)}{\sqrt{3}}, \quad x_{17} = \frac{1}{2} \left(Q(1 - 146Q^2) x_1 + (1 + 96Q^2) x_8 + 6 x_{12}\right), \\
    x_{18} &= -10 Q^3 x_1 + \frac{1}{3}x_8 + x_{12}, \quad x_{19} = \frac{Q \left(Q(66Q^2-1)x_1 - 3 x_8 - 12 x_{12}\right)}{\sqrt{3}}.
\end{aligned}
\end{equation}
Substituting these relations into the ansatz and rearranging the terms, we obtain
\begin{equation}
    W_5 = x_1 \widehat W_5 + x_8(W_2\partial{W_2}) + x_{12}\partial^{3}{W_2},
\end{equation}
where $W_5$ is decomposed into terms constructed solely from the lower-spin current $W_2$ and the remaining contribution that cannot be expressed only in terms of $W_2$. The latter, denoted by $\widehat{W}_5$, is given by
\begin{align}
    \widehat W_5
    &= r_5 + \frac{1}{\sqrt{3}}(p w_4) - Q\partial{w_4} - \frac{1}{2 \sqrt{3}}(p(w_2 w_2)) - \frac{1}{6 \sqrt{3}}(p(p(p w_2))) + \frac{10 Q^2}{\sqrt{3}}(\partial^{2}{p} w_2) \nonumber\\
    &\quad + 3 \sqrt{3} Q^2 (\partial{p}\partial{w_2}) + \frac{1-24 Q^2}{4 \sqrt{3}}(p \partial^{2}{w_2}) + \frac{1}{24 \sqrt{3}}(p(p(p(p p)))) - \frac{4 Q}{3}(p(p(p \partial{p}))) \nonumber\\
    &\quad + \frac{13 \sqrt{3} Q^2}{2}(p(\partial{p} \partial{p})) + \frac{11 Q^2}{\sqrt{3}}(p(p \partial^{2}{p})) + \frac{Q}{2}(1-146 Q^2) (\partial{p} \partial^{2}{p}) - 10 Q^3(p \partial^{3}{p}) \nonumber\\
    &\quad - \frac{Q^2 (1-66 Q^2)}{\sqrt{3}}\partial^{4}{p}.
\end{align}
The coefficient $x_1$ fixes the overall normalization of $\widehat W_5$, while $x_8$ and $x_{12}$ are the coefficients of the terms constructed from $W_2$. Requiring $W_5$ to be primary determines $x_8$ and $x_{12}$ as follows:
\begin{equation}
    x_8 = Qx_1, \quad x_{12} = -\frac{1}{3}Q(1-15Q^2)x_1.
\end{equation}
Using these relations and fixing the normalization as $x_1=1$, the primary spin-$5$ current takes the form
\begin{equation}\label{eq:primW5E6}
    \widetilde{W}_5 = \widehat{W}_5 + Q(W_2 \partial{W_2}) - \frac{Q}{3}(1-15Q^2)\partial^{3}{W_2}.
\end{equation}

The higher-spin currents $W_s~(s=6,8,9,12)$ can be constructed similarly, but the number of terms in the ansatz grows rapidly with the spin. It is more efficient to obtain them from the OPEs of the lower-spin currents \cite{Keller:2011ek}:
\begin{equation}\label{eq:WE6currents}
    W_6 = \frac{1}{1-12Q^2}\{\widetilde{W}_5 \widetilde{W}_5\}_4, \quad W_8 = \{\widetilde{W}_5 \widetilde{W}_5\}_2, \quad W_9 = \{\widetilde{W}_5 W_6\}_2, \quad W_{12} = \{W_6 W_8\}_2.
\end{equation}
These currents are not primary in general, and the corresponding primary currents $\widetilde{W}_s$ are obtained by adding the terms built from the $\W$-currents of lower spin, with coefficients fixed by requiring them to be primary. As a consistency check, we have also solved the most general spin-$6$ ansatz subject to the zero commutation relation associated with $\alpha_5$. The coefficients left undetermined by the relation can be chosen so that the solution coincides with $\{\widetilde{W}_5 \widetilde{W}_5\}_4$.

The $\W E_6$ algebra has previously been constructed from the $\W A_5$ algebra in \cite{Keller:2011ek}. The construction differs from the present one in the choice of the node removed from the Dynkin diagram of $E_6$. Both constructions realize the algebra on six free bosons, and the two bases of the root space are related by an orthogonal transformation. Under this map, our $W_2$ coincides with the energy-momentum tensor of \cite{Keller:2011ek}, and our primary current $\widetilde{W}_5$ coincides with their spin-$5$ generator up to an overall normalization.\footnote{The conventions of \cite{Keller:2011ek} differ from ours. Denoting their parameters by $b'$ and $Q'$, their screening currents are  $:\exp({b'}^{\pm1}\alpha_i\cdot\phi):$ with $Q' = b' + {b'}^{-1}$, while ours are given in \eqref{eq:scr}. 
The two are related by $b = ib'$ and $Q = iQ'$.} This confirms that the two constructions give the same algebra.

\subsection{\texorpdfstring{$\mathcal{W}$}{W}-charges}\label{sec:WchargeE6}

Having constructed the $\W$-currents, we now turn to the representation theory of the $\W E_6$ algebra. A highest-weight representation of the $\W\g$ algebra is characterized by the central charge together with the eigenvalues of the zero modes of the $\W$-currents, which we refer to as the $\W$-charges. In the free field realization, this representation is generated from a vertex operator labeled by a momentum $\lambda$, so that computing the $\W$-charges amounts to giving the dictionary between the free field momentum and the data labeling the representation. This is the starting point for studying the module structure of the algebra.

Let us now give the definition. The vertex operator is $V_{\lambda}(z) = ~:e^{i\lambda\cdot\phi}:$, and the corresponding state is $\ket{\lambda}$. Expanding the currents in modes as $W_s(z)=\sum_n(W_s)_n z^{-n-s}$, this state is annihilated by the positive modes and is an eigenstate of the zero modes,
\begin{equation}
    (W_s)_n\ket{\lambda}=0 \;\; (n>0), \quad (W_s)_0\ket{\lambda}=\Delta_s\ket{\lambda}.
\end{equation}
Acting on $\ket{\lambda}$ with the negative modes generates a highest-weight module. Equivalently, the $\W$-charge $\Delta_s$ is read off from the leading singularity of the OPE with the vertex operator:
\begin{equation}\label{eq:WchargeOPE}
    W_s(z)V_{\lambda}(w) \sim \frac{\Delta_{s}V_{\lambda}(w)}{(z-w)^{s}} + \text{lower order terms}.
\end{equation}
The $\W$-charge $\Delta_2$ is the conformal weight of the vertex operator $V_{\lambda}$.

Using the OPE \eqref{eq:WchargeOPE}, we can compute the $\W$-charges for the generators $W_{2}$, $\widetilde{W}_5$, $W_6$, $W_8$, $W_9$, and $W_{12}$ of the $\W E_6$ algebra in \eqref{eq:EMtensorWE6}, \eqref{eq:primW5E6}, and \eqref{eq:WE6currents}. We express them in terms of the Casimir invariants of $E_6$. The Casimir invariants are evaluated in the $27$-dimensional fundamental representation of $E_6$, whose weights are $\mu_a~(a=1,\ldots,27)$. We denote by $H$ the Cartan generator in this representation, whose eigenvalues on the weight vectors $\ket{\mu_a}~(a=1,\ldots,27)$ are the weights $\mu_a$: $H\ket{\mu_a} = \mu_a\ket{\mu_a}$. The Cartan generators $H_i = \alpha_i(E_6) \cdot H$ are diagonal in this basis:
\begin{equation}
    H_i = \sum_{a=1}^{27} (\mu_a \cdot \alpha_i(E_6)) E_{aa}, \quad (i=1,\ldots,6),
\end{equation}
where $E_{ab}$ are the $27 \times 27$ matrix units: $(E_{ab})_{cd} = \delta_{ac}\delta_{bd}$. Their explicit forms are shown in Appendix~\ref{app:Lie}. Defining the shifted momentum $\widetilde{\lambda} = \lambda + Q\rho(E_6)$, we introduce the Casimir invariants
\begin{equation}
    C_{k} =\Tr_{\mathbf{27}}(\widetilde{\lambda} \cdot H)^{k}, \quad (k = 2,5,6,8,9,12).
\end{equation}
The $\W$-charges are expressed in terms of these invariants as
\begin{align}
    \Delta_2 &= \frac{C_2}{12} - 39Q^2 = \frac{C_2}{12} + \frac{c(E_6)-6}{24}, \\
    \Delta_5 &= -\frac{C_5}{60}, \\
    \Delta_6 &= \frac{C_6}{9} - \frac{C_2^3}{1296} + \frac{5 C_2^2}{216}(2-91 Q^2) + \frac{5 C_2}{27}(2 - 444 Q^2 + 15309 Q^4)
    \nonumber\\
    &\quad - \frac{130Q^2}{3}(4 - 654 Q^2 + 20463 Q^4),
\end{align}
{\allowdisplaybreaks\small
\begin{align}
    \Delta_8 &= \frac{C_8}{60} - \frac{C_6 C_2}{270} + \frac{2 C_6}{9}(3-37 Q^2) + \frac{C_2^4}{62208} - \frac{C_2^3}{648}(3-37 Q^2) + \frac{C_2^2}{324}(25 - 1665 Q^2 + 16401 Q^4) \nonumber\\
    &\quad  + \frac{10 C_2}{27}(1 - 294 Q^2 + 14031 Q^4 - 127773 Q^6) - \frac{52 Q^2}{3}(10 - 1965 Q^2 + 82755 Q^4 - 727383 Q^6), \\
    \Delta_9 &= -\frac{2 C_9}{21} + \frac{C_5 C_2^2}{216} - \frac{C_5 C_2}{27}(37-753 Q^2) - \frac{2 C_5}{135}(7190 - 404760 Q^2 + 5006637 Q^4), \\
    \Delta_{12} &= \frac{16 C_{12}}{135} - \frac{11 C_8 C_2^2}{1215} + \frac{C_8 C_2}{81} (433-2796 Q^2) + \frac{4C_8}{135} (56825 - 1046187 Q^2 + 4746582 Q^4) \nonumber\\
    &\quad - \frac{8 C_6^2}{1215} + \frac{17 C_6 C_2^3}{10935} - \frac{C_6 C_2^2}{729}(725-3888 Q^2) - \frac{4 C_6 C_2}{1215}(56700 - 1017499 Q^2 + 4915824 Q^4) \nonumber\\
    &\quad + \frac{8 C_6}{81} (232695 - 9282410 Q^2 + 115557156 Q^4 - 453615912 Q^6) + \frac{C_5^2}{135}(521-3873 Q^2) \nonumber\\
    &\quad - \frac{209 C_5^2 C_2}{24300} - \frac{C_2^6}{174960} + \frac{C_2^5}{419904}(1601-7164 Q^2) + \frac{C_2^4}{34992}(17715 - 544457 Q^2 + 4454118 Q^4) \nonumber\\
    &\quad - \frac{C_2^3}{4374}(433460 - 5314830 Q^2 - 189696477 Q^4 + 2287669284 Q^6) \nonumber\\
    &\quad + \frac{C_2^2}{729}(2664500 - 365347425 Q^2 + 16327732650 Q^4 - 292597319538 Q^6 + 1736070778440 Q^8) \nonumber\\
    &\quad + \frac{40C_2}{243}(126170 - 41960580 Q^2 + 3858594030 Q^4 - 146307615390 Q^6 \nonumber\\
    &\qquad\qquad\quad + 2402201707191 Q^8 - 13587087836772 Q^{10}) \nonumber\\
    &\quad - \frac{16Q^2}{27} (16402100 - 4103973900 Q^2 + 334292525775 Q^4 - 11881307876100 Q^6 \nonumber\\
    &\quad\qquad\qquad + 187408555866792 Q^8 - 1034633243838600 Q^{10}).
\end{align}}
Each $\Delta_s$ is a polynomial in the Casimir invariants whose leading term is $C_s$. The results exhibit a definite parity under the $\mathbb{Z}_2$ outer-automorphism of $E_6$ that interchanges $\alpha_1$ and $\alpha_5$, and $\alpha_2$ and $\alpha_4$. With the choice of $\W$-currents as in \cite{Keller:2011ek}, the $\mathbb{Z}_2$ action reads $W_s \to (-1)^s W_s$ and $C_k \to (-1)^k C_k$. It follows that $\Delta_s\to(-1)^s\Delta_s$ under the $\mathbb{Z}_2$ action. In particular, we observe that the terms $C_5$ in $\Delta_6$ and $C_5C_2$ in $\Delta_8$, which are allowed by degree counting, do not appear.

\section{Free Field Realization of \texorpdfstring{$\mathcal{W}E_7$}{WE7} Algebra}\label{sec:WE7}

The $\W E_7$ algebra is generated by the $\W$-currents $W_{s}$ of spins $s=2,6,8,10,12,14$, and $18$. Following the recursive construction described in Section~\ref{sec:Wg}, we discuss two constructions of the $\W E_7$ algebra based on different rank-$6$ subalgebras of $E_7$.\footnote{It would be possible to construct the $\W E_7$ algebra from the $\W A_6$ algebra. This needs a large number of basis operators to express the $\W$-currents. We do not try this approach.} In Section~\ref{sec:WE7fromWD6}, we construct the $\W E_7$ algebra from the $\W D_6$ algebra and an additional free boson, following the same procedure as for the $\W E_6$ algebra in Section~\ref{sec:WE6}. In Section~\ref{sec:WE7fromWE6}, we give an alternative construction based on the $\W E_6$ algebra obtained in Section~\ref{sec:WE6} and an additional free boson. We show that the two constructions are related by a change of basis of the free fields, providing a consistency check of the recursive construction.

\subsection{Construction from \texorpdfstring{$\mathcal{W}D_6$}{WD6} Algebra}\label{sec:WE7fromWD6}

As in the case of the $\W E_6$ algebra, we construct the $\W E_7$ algebra from the $\W D_6$ algebra. The simple roots of $E_7$ are expressed in terms of the simple roots of $D_6$ and its orthogonal direction as:
\begin{equation}\label{eq:E7rootsD6}
    \alpha_{i}(E_7) = \alpha_{i}(D_6), \;\; (i=1,\ldots,5), \quad
    \alpha_{6}(E_7) = -\Lambda_5(D_6) + \frac{1}{\sqrt{2}}e_{7}, \quad
    \alpha_{7}(E_7) = \alpha_{6}(D_6),
\end{equation}
where $e_7$ is a unit vector orthogonal to the root space of $D_6$. For this choice of the simple roots, the Weyl vector of $E_7$ is written as
\begin{equation}
    \rho(E_7) = \rho(D_6) + \frac{17}{\sqrt{2}}e_7,
\end{equation}
and the energy-momentum tensor of the $\W E_7$ algebra is obtained as
\begin{equation}
\label{eq:WE7w2}
    W_2 = w_2 + \frac{1}{2}(pp) - \frac{17Q}{\sqrt{2}}\partial{p},
\end{equation}
where $p = i\partial\varphi(z)$ and $\varphi(z)$ is the additional free boson. The central charge is $c(E_7) = 7 - 2394 Q^2$. As in the previous section, the generators $w_s~(s=2,4,6,8,10)$ and $r_6$ of the $\W D_6$ algebra and $p$ satisfy the zero commutation relations with $Q^{\pm}_i~(i=1,2,3,4,5,7)$. Therefore, it is sufficient to impose the relation associated with $\alpha_6$. For the currents of spins $3, 4$, and $5$, we examine the most general ansatz and impose the zero commutation relation associated with $\alpha_6$. These currents are shown to be expressed only in terms of $W_2$.

We next discuss the spin-$6$ current. The most general ansatz for $W_6$ is written in terms of the generators $w_s~(s=2,4,6,8,10)$ and $r_6$ of the $\W D_6$ algebra and $p$ as
\begin{align}
    \nonumber
    W_6&=x_1 r_6 + x_2 w_6 + x_3(w_2 w_4) + x_4(p(p w_4)) + x_5(\partial{p} w_4) + x_6(p \partial{w_4}) + x_7\partial^{2}{w_4}+x_8(w_2(w_2 w_2))\\
    \nonumber
    &\quad +x_9(p(p(w_2 w_2))) + x_{10}(\partial{p}(w_2 w_2)) + x_{11}(p(w_2 \partial{w_2})) + x_{12}(\partial{w_2} \partial{w_2}) + x_{13}(w_2 \partial^{2}{w_2})\\
    \nonumber
    &\quad +x_{14}(p(p(p(p w_2)))) + x_{15}(p(p(\partial{p} w_2))) + x_{16}(p(p(p \partial{w_2}))) + x_{17}(\partial{p}(\partial{p} w_2))\\
    \nonumber
    &\quad +x_{18}(p(\partial^{2}{p}w_2))+x_{19}(p(\partial{p} \partial w_2)) + x_{20}(p(p \partial^{2}{w_2})) + x_{21}(\partial^{3}{p} w_2) + x_{22}(\partial^{2}{p} \partial{w_2}) \\
    \nonumber
    &\quad +x_{23}(\partial p\partial^{2}{w_2})+x_{24}(p\partial^{3}{w_2})+ x_{25}\partial^{4}{w_2} + x_{26}(p(p(p(p(p p))))) + x_{27}(p(p(p(p \partial{p}))))\\
    \nonumber
    &\quad +x_{28}(p(p(p\partial^{2}{p})))+x_{29}(p(p(\partial{p}\partial{p})))+x_{30}(p(p \partial^{3}{p}))+x_{31}(p(\partial{p} \partial^{2}{p}))+x_{32}(p \partial^{4}{p})\\
    &\quad +x_{33}(\partial{p}(\partial{p} \partial{p}))+x_{34}(\partial{p} \partial^{3}{p})+x_{35}(\partial^{2}{p} \partial^{2}{p})+x_{36}\partial^{5}{p}.
\end{align}
Imposing the zero commutation relation associated with $\alpha_6$ determines the coefficients $x_j$ in terms of $x_1$, $x_8$, $x_{12}$, $x_{13}$, and $x_{25}$. Rearranging the terms, $W_6$ can be written as
\begin{equation}
    W_6 = x_1 \widehat{W}_6 + x_8(W_2(W_2 W_2)) + x_{12}(\partial W_2 \partial W_2) + x_{13}(W_2 \partial^{2}W_2) + x_{25}\partial^{4}W_2,
\end{equation}
where $W_6$ is decomposed into terms constructed solely from the lower-spin current $W_2$ and the remaining contribution that cannot be expressed solely in terms of $W_2$. The latter, denoted by $\widehat{W}_6$, is given by

{\small
\begin{align}\label{eq:W6 of WE7}
    \widehat{W}_6
    &= r_6 + \frac{w_6}{5} - \frac{1}{15}(w_2 w_4) - \frac{1}{3}(p(p w_4)) + \frac{13\sqrt{2}Q}{15}(\partial{p} w_4)+\frac{6\sqrt{2}Q}{5}(p \partial{w_4}) + \frac{1 - 96 Q^2}{30}\partial^{2}{w_4}
    \nonumber\\
    &\quad +\frac{11}{60}(p(p(w_2 w_2))) - \frac{11Q}{30\sqrt{2}}(\partial{p}(w_2 w_2)) - \frac{22\sqrt{2}Q}{15}(p(w_2 \partial{w_2})) + \frac{1}{120}(p(p(p(p w_2))))
    \nonumber\\
    &\quad -\frac{Q}{30\sqrt{2}}(p(p(\partial{p}w_2)))-\frac{2\sqrt{2}Q}{15}(p(p(p \partial{w_2}))) - \frac{1 + 463 Q^2}{60}(\partial{p}(\partial{p} w_2))-\frac{28Q^2}{15}(p(\partial^{2}{p}w_2))
    \nonumber\\
    &\quad +\frac{1-86 Q^2}{30}(p(\partial{p} \partial{w_2}))-\frac{1}{120}(11 - 192 Q^2) (p (p \partial^{2}{w_2})) + \frac{Q(1 + 1088Q^2)}{20\sqrt{2}}(\partial^{3}{p} w_2)
    \nonumber\\
    &\quad -\frac{Q(13 - 2550 Q^2)}{30\sqrt{2}}(\partial^{2}{p} \partial{w_2}) + \frac{Q(19 + 2352Q^2)}{60\sqrt{2}}(\partial{p} \partial^{2}{w_2}) + \frac{\sqrt{2}Q}{15}(7 - 150 Q^2)(p \partial^{3}{w_2})
    \nonumber\\
    &\quad -\frac{1}{240}  (p(p(p(p(p p))))) + \frac{17Q}{40\sqrt{2}}(p(p(p(p \partial{p})))) + \frac{1}{120}(1-1159 Q^2)(p(p(\partial{p}\partial{p})))
    \nonumber\\
    &\quad -\frac{44Q^2}{15}(p(p(p \partial^{2}{p})))-\frac{Q(3 - 5617Q^2)}{60\sqrt{2}}(\partial{p}(\partial{p} \partial{p})) - \frac{Q(7 - 5122Q^2)}{30\sqrt{2}}(p(\partial{p} \partial^{2}{p}))
    \nonumber\\
    &\quad -\frac{Q(1 - 4224 Q^2)}{120\sqrt{2}}(p(p\partial^{3}{p}))-\frac{1}{240}(1 - 704Q^2 + 108420Q^4) (\partial^{2}{p} \partial^{2}{p})
    \nonumber\\
    &\quad +\frac{1}{360}(1+293Q^2 - 208224Q^4)(\partial{p} \partial^{3}{p}) - 28Q^4(p \partial^{4}{p}) - \frac{17(59 - 5760 Q^2)Q^3}{300 \sqrt{2}}\partial^{5}{p}.
\end{align}}
Other higher-spin generators are obtained from the OPEs as
\begin{equation}\label{eq:WE7currents}
\begin{aligned}
    W_8 &= \{\widehat{W}_6 \widehat{W}_6\}_4, & W_{10} &= \{\widehat{W}_6 \widehat{W}_6\}_2, & W_{12} &= \{\widehat{W}_6 W_8\}_2, \\
    W_{14} &= \{\widehat{W}_6 W_{10}\}_2, & W_{18} &= \{\widehat{W}_6 W_{14}\}_2.
\end{aligned}
\end{equation}

\subsection{Construction from \texorpdfstring{$\mathcal{W}E_6$}{WE6} Algebra}\label{sec:WE7fromWE6}

We can also construct the $\W E_7$ algebra from the $\W E_6$ algebra. We decompose the simple roots of $E_7$ in terms of those of $E_6$ and its orthogonal direction by $e_7'$ as
\begin{equation}\label{eq:E7rootsE6}
    \alpha_1(E_7) = -\Lambda_1(E_6) + \sqrt{\frac{2}{3}}e_7', \quad \alpha_i(E_7) = \alpha_{i-1}(E_6)~~(i=2,\ldots,7),
\end{equation}
where $\Lambda_1(E_6)$ is the fundamental weight of $E_6$ dual to the simple root $\alpha_1(E_6)$. The additional simple root $\alpha_\ast$ is $\alpha_1(E_7)$.
The Weyl vector of $E_7$ can be expressed as
\begin{equation}
    \rho(E_7) = \rho(E_6) + 9\sqrt{\frac{3}{2}}e_7'.
\end{equation}
Using this decomposition of the Weyl vector, the energy-momentum tensor is obtained as
\begin{equation}\label{eq:WE7w22}
    W_2 = w_2 + \frac{1}{2}(pp) - 9\sqrt{\frac{3}{2}}Q\partial p,
\end{equation}
where $w_2$ is now the generator of the $\W E_6$ algebra given in \eqref{eq:EMtensorWE6}, and $p$ corresponds to the $e_7'$ direction.

We can show that the above $W_2$ coincides with that obtained from the $\W D_6$ algebra. In the decompositions of $E_7$ \eqref{eq:E7rootsD6} and \eqref{eq:E7rootsE6}, the simple roots of $D_6$ and $E_6$ can be decomposed into those of a common $D_5$ and an orthogonal direction. The decomposition of $D_6$ is given by \eqref{eq:Drdecom} with $r=6$, and that of $E_6$ by \eqref{eq:E6root}. Both realizations are therefore built on the generators of the $\W D_5$ algebra and two additional free bosons. In the $\W D_6$-based construction the two bosons are $p_1$ along $e_1$ and $p_7$ along $e_7$ in \eqref{eq:Drdecom}, \eqref{eq:E7rootsD6}. In the $\W E_6$-based construction they are $p_6'$ along $e_6$ and $p_7'$ along $e_7'$ in \eqref{eq:E6root}, \eqref{eq:E7rootsE6}. The two decompositions are related by an orthogonal transformation that leaves the $\W D_5$ generators unchanged and mixes the two remaining bosons as
\begin{equation}\label{eq:changebasisWE7}
    \mqty(p_6' \\ p_7') =
    \frac{1}{\sqrt{3}}\mqty(
        -1 & \sqrt{2} \\
        \sqrt{2} & 1
    )\mqty(p_1 \\ p_7).
\end{equation}
For the spin-$2$ current, \eqref{eq:WE7w2} and \eqref{eq:WE7w22} are then identical under this transformation.

For the spin-$6$ current, the most general ansatz can be written as
{\small
\begin{align}
    U_6 &= x_1 w_6 + x_2 (p w_5) + x_3 \partial{w_5} + x_4 (w_2(w_2 w_2)) + x_5 (p(p(w_2 w_2))) + x_6 (\partial{p}(w_2 w_2))
    \nonumber\\
    &\quad + x_7(p(w_2 \partial{w_2})) + x_8(\partial{w_2}\partial{w_2}) + x_9 (w_2 \partial^2{w_2}) + x_{10} (p(p(p(p w_2)))) + x_{11} (p(p(\partial{p} w_2)))
    \nonumber\\
    &\quad + x_{12} (p(p(p\partial{w_2})))+x_{13}(\partial{p}(\partial{p} w_2)) + x_{14} (p(\partial^2{p} w_2)) + x_{15} (p(\partial{p} \partial{w_2})) + x_{16} (p(p \partial^2{w_2}))
    \nonumber\\
    &\quad + x_{17} (\partial^3{p} w_2) + x_{18} (\partial^2{p} \partial{w_2}) + x_{19}(\partial{p} \partial^2{w_2}) + x_{20} (p \partial^3{w_2}) + x_{21} \partial^4{w_2} + x_{22}(p(p(p(p(p p)))))
    \nonumber\\
    &\quad + x_{23} (p(p(p(p \partial{p})))) + x_{24}(p(p(\partial{p} \partial{p}))) + x_{25}(p(p(p\partial^2{p}))) + x_{26} (\partial{p}(\partial{p} \partial{p}))
    \nonumber\\
    &\quad + x_{27}(p(\partial{p} \partial^2{p})) + x_{28}(p(p\partial^3{p})) + x_{29}(\partial^2{p} \partial^2{p}) + x_{30}(\partial{p} \partial^3{p}) + x_{31} (p \partial^4{p}) + x_{32} \partial^5{p},
\end{align}}
where $w_5$ and $w_6$ are the spin-$5$ and spin-$6$ currents of the $\W E_6$ algebra, given by $\widetilde{W}_5$ in \eqref{eq:primW5E6} and $W_6$ in \eqref{eq:WE6currents}. The zero commutation relation associated with $\alpha_1(E_7)$ determines the coefficients $x_j$ in terms of $x_1,x_4,x_8,x_9$, and $x_{21}$. We obtain
\begin{equation}
    U_6 = x_1 \widehat{U}_6 + x_4 (W_2(W_2 W_2)) + x_8(\partial{W_2} \partial{W_2}) + x_9(W_2 \partial^{2}{W_2}) + x_{21} \partial^{4}{W_2},
\end{equation}
where $U_6$ is decomposed into terms constructed solely from the lower-spin current $W_2$ and the remaining contribution that cannot be expressed only in terms of $W_2$. $\widehat{U}_6$ is given by
{\allowdisplaybreaks\small
\begin{align}
    \widehat{U}_6
    &= w_6 - 20\sqrt{\frac{2}{3}} (p w_5) + 36Q \partial{w_5} + \frac{4}{3} (p(p(w_2 w_2))) - 20\sqrt{\frac{2}{3}}Q (\partial{p}(w_2 w_2)) - 8\sqrt{\frac{2}{3}}Q (p(w_2 \partial{w_2})) \nonumber\\
    &\quad - \frac{4}{9} (p(p(p(p w_2)))) + 8\sqrt{\frac{2}{3}}Q (p(p(\partial{p} w_2))) + 8\sqrt{\frac{2}{3}}Q (p(p(p \partial{w_2}))) - \frac{2}{9}(1 - 330Q^2)(\partial{p}(\partial{p} w_2)) \nonumber\\
    &\quad + \frac{2}{9}(5 - 438Q^2)(p(\partial^2{p} w_2)) + \frac{4}{3}(1 - 114Q^2)(p(\partial{p} \partial{w_2})) - \frac{1}{9}(1 + 390Q^2)(p(p \partial^2{w_2})) \nonumber\\
    &\quad - \frac{\sqrt{2}Q(29-1710Q^2)}{3\sqrt{3}} (\partial^3{p} w_2) - 16\sqrt{\frac{2}{3}}Q(1-63Q^2)(\partial^2{p} \partial{w_2}) - \sqrt{6}Q(1-210Q^2)(\partial{p} \partial^2{w_2}) \nonumber\\
    &\quad + 2\sqrt{\frac{2}{3}}Q(1+30Q^2)(p \partial^3{w_2}) + \frac{2}{9}(p(p(p(p(p p))))) - 6\sqrt{6}Q(p(p(p(p \partial{p})))) \nonumber\\
    &\quad + \frac{5}{9}(1+318Q^2)(p(p(\partial{p} \partial{p}))) + \frac{1}{9}(5+426Q^2)(p(p(p \partial^2{p}))) - \sqrt{6}Q(1+150Q^2)(\partial{p}(\partial{p} \partial{p})) \nonumber\\
    &\quad - 3\sqrt{6}Q(3-26Q^2)(p(\partial{p} \partial^2{p})) - \frac{Q(11+30Q^2)}{\sqrt{6}}(p(p \partial^3{p})) + 7Q^2(9-382Q^2)(\partial^2{p} \partial^2{p}) \nonumber\\
    &\quad + \frac{1}{3}Q^2(229-8670Q^2)(\partial{p} \partial^3{p}) + \frac{5}{54}(1-60Q^2+5508Q^4)(p \partial^4{p}) - \frac{Q(1-105Q^2+4410Q^4)}{\sqrt{6}}\partial^5{p}.
\end{align}
}
Under the change of basis \eqref{eq:changebasisWE7}, $\widehat{U}_6$ is related to $\widehat{W}_6$ as
\begin{equation}
\begin{aligned}
    \widehat{U}_6 &= 20\widehat{W}_6 + \frac{2}{3}(W_2(W_2 W_2)) + \frac{1}{3}(1 + 66 Q^2)(\partial W_2 \partial W_2) + \frac{1}{9}(1 + 276 Q^2)(W_2\partial^2 W_2) \\
    &\quad - \frac{1}{108}(5 + 1626 Q^2 - 55080 Q^4)\partial^4 W_2.
\end{aligned}
\end{equation}
Here, $\widehat{W}_6$ denotes the part of the spin-$6$ current in the $\W D_6$-based construction that cannot be expressed in terms of $W_2$.

\subsection{\texorpdfstring{$\mathcal{W}$}{W}-charges}

The $\W$-charges of the $\W E_7$ algebra are defined as in Section~\ref{sec:WchargeE6}, now with the shifted momentum $\widetilde{\lambda} = \lambda + Q\rho(E_7)$. All the Casimir invariants of $E_7$ have even degrees: $k=2,6,8,10,12,14,18$. We evaluate the Casimir invariants in the $56$-dimensional fundamental representation as $C_k=\mathrm{Tr}_{\mathbf{56}}(\widetilde{\lambda}\cdot H)^k$, with the Cartan generators $H_i$ given in Appendix~\ref{app:Lie}. 

The $\W$-charges for $W_2$, $\widehat{W}_6$, $W_8$, and $W_{10}$ in \eqref{eq:WE7w2}, \eqref{eq:W6 of WE7}, and \eqref{eq:WE7currents} are expressed in terms of the Casimir invariants:
{\allowdisplaybreaks\small
\begin{align}
    \Delta_2 &= \frac{C_2}{24} - \frac{399Q^2}{4} = \frac{C_2}{24} + \frac{c(E_7)-7}{24},
    \\
    \Delta_6 &= \frac{C_6}{360} - \frac{C_2^3}{138240} + \frac{C_2^2}{69120}(4 - 3227 Q^2) + \frac{Q^2 C_2}{1280}(904 + 187421 Q^2)
    \nonumber\\
    &\quad -\frac{19 Q^4}{640}(68124 + 2371345 Q^2),
    \\
    \Delta_8 &= \frac{C_8}{1800}(25 - 414 Q^2) - \frac{C_6C_2}{21600}(45 - 76 Q^2) + \frac{C_6}{10800}(3690 - 637037 Q^2 - 4683168 Q^4)
    \nonumber\\
    &\quad +\frac{C_2^4}{199065600}(597 + 5122Q^2) - \frac{C_2^3}{24883200}(12102 - 3191119 Q^2 + 18656778 Q^4)
    \nonumber\\
    &\quad + \frac{C_2^2}{8294400}(189672 - 16111812 Q^2 - 906418003 Q^4- 670280773086 Q^6)
    \nonumber\\
    &\quad +\frac{C_2}{230400}(50896 - 127763464 Q^2 + 20458876818 Q^4 + 774316526481 Q^6 - 12223895849646 Q^8)
    \nonumber\\
    &\quad -\frac{19Q^2}{153600}(4275264 - 9672623184Q^2 + 1660798850904Q^4 + 59377789779377Q^6
    \nonumber\\
    &\qquad\qquad\qquad - 4963165941139686Q^8), \\
    \Delta_{10} &= \frac{C_{10}}{1050} - \frac{C_8C_2}{7560} + \frac{C_8}{1800}(250-4209 Q^2) + \frac{C_6C_2^2}{518400} - \frac{C_6C_2}{129600}(2688 + 5045Q^2)
    \nonumber\\
    &\quad + \frac{C_6}{43200}(39600 - 7598480Q^2 - 110085297Q^4) + \frac{7C_2^5}{1194393600}
    \nonumber\\
    &\quad + \frac{C_2^4}{597196800}(17766 + 272393Q^2) - \frac{C_2^3}{49766400}(69216 - 20330492Q^2 + 163708145Q^4)
    \nonumber\\
    &\quad + \frac{C_2^2}{8294400}(237048 - 8578272Q^2 - 5415396430Q^4 - 868296663903Q^6)
    \nonumber\\
    &\quad + \frac{C_2}{921600}(212864 + 799422816Q^2 + 140060942464Q^4 + 3328166516136Q^6 + 31882827019535Q^8) \nonumber\\
    &\quad - \frac{19 Q^2}{153600}(4470144 - 15463729008Q^2 + 2970360251328Q^4 - 41120835586970Q^6
    \nonumber\\
    &\qquad\qquad\qquad -5177685668783565Q^8).
\end{align}}
Each $\Delta_s$ is a polynomial in the Casimir invariants whose leading term is $C_s$. The $\W$-charge for the spin-$12$ generator $W_{12}$ in \eqref{eq:WE7currents} is given in Appendix~\ref{app:WE7Charge}. We have also constructed $W_{14}$ and $W_{18}$ from \eqref{eq:WE7currents}, but the computation of their $\W$-charges is left for future work.

\section{Free Field Realization of \texorpdfstring{$\mathcal{W}E_8$}{WE8} Algebra}\label{sec:WE8}

The $\W E_8$ algebra contains the generators $W_s$ with $s=2$, $8$, $12$, $14$, $18$, $20$, $24$, and $30$, where the spins correspond to the orders of Casimirs of $E_8$. The $\W$-currents may be constructed from the $\W E_7$ algebra and a free boson. Instead, we construct them based on the $\W D_7$ algebra and a free boson. Under the decomposition of the root space of $E_8$ into that of $D_7$ and its orthogonal direction, the simple roots of $E_8$ are expressed as
\begin{align}
    \alpha_i(E_8) &= \alpha_i(D_7), ~~ (i=1,\ldots,5),  \quad
    \alpha_6(E_8) = \alpha_7(D_7), \nonumber\\
    \alpha_7(E_8) &= -\Lambda_7(D_7) + \frac{1}{2}e_8, \quad
    \alpha_8(E_8) = \alpha_6(D_7),
\end{align}
where $e_8$ is the unit vector orthogonal to the root space of $D_7$. The Weyl vector of $E_8$ is 
\begin{equation}
    \rho(E_8) = \rho(D_7)+23 e_8.
\end{equation}

We now construct the spin-8 current.
Let $w_2,w_4,w_6,w_8,w_{10},w_{12}$, and $r_7$ be the generators of the $\W D_7$ algebra.
We introduce a free boson $\varphi$ and $p=i\partial \varphi$.
The spin-2 current of the $\W E_8$ algebra is 
\begin{equation}
    W_2 = w_2+\frac{1}{2}(pp)-23Q\partial p.
\end{equation}
The central charge is $c(E_8)=8-7440 Q^2$.
For the spin-8 current, we take the general ansatz such that the current is expressed in terms of the $\W$-currents of $\W D_7$ and a free boson $\varphi$ together with their derivatives, which includes 104 terms. This ansatz is consistent with the zero commutation relation with the screening charges $Q^{\pm}_i$ for $i=1,\dots,7$ automatically. The zero commutation relation with $Q_8^{\pm}$ leads to constraints on these 104 coefficients and can be solved with 8 free parameters.
Then, the $\W$-current is expressed as
\begin{align}
    W_8
    &=
    x_1 \widehat{W}_8+x_{2}(W_2(W_2(W_2W_2)))+x_{3} (W_2 (W_2\partial^2 W_2))+x_{4}(W_2(\partial W_2\partial W_2))
    \nonumber\\
    &\quad
    +x_{5} (W_2\partial^4 W_2)
    +x_{6}(\partial W_2 \partial^3 W_2)+x_{7} (\partial^2 W_2 \partial^2 W_2)+x_{8}\partial^6 W_2.
\end{align}
Here, $\widehat{W}_8$ is a non-trivial part of the current, which is found to be
{\allowdisplaybreaks\small
\begin{align}
    \widehat{W}_8&
    =
    w_8
    -7 (p r_7) 
    +23Q \partial r_7
    -\frac{3}{5} (w_2 w_6)
    + \frac{7}{10} (p(p w_6))
    -5Q (p\partial w_6)
    -\frac{11}{5}Q (\partial p w_6)
    +\left(4Q^2+\frac{1}{20}\right) \partial^2 w_6
    \nonumber\\
    &
    +\frac{1}{6}(w_4 w_4)
    +\frac{1}{30}(w_2 (w_2 w_4))
    +\left(3 Q^2-\frac{1}{5}\right)(w_2 \partial^2 w_4)
    +\left(-8 Q^2-\frac{1}{30}\right)(\partial w_2 \partial w_4)
    +\left(-8Q^2-\frac{11}{60}\right)(\partial^2 w_2  w_4)
    \nonumber\\
    &-\frac{7}{30}(p(p(w_2 w_4)))
    +\frac{7 Q}{5}(p(w_2\partial w_4))
    +\frac{7 Q}{3}(p(\partial w_2 w_4))
    +\frac{7 Q}{15}(\partial p( w_2 w_4))
    -\frac{7}{24}(p(p(p(p w_4))))
    \nonumber\\
    &+\frac{47 Q}{10}(p(p(p \partial w_4)))
    +\frac{241Q}{30}(p(p(\partial p  w_4)))
    +\left(\frac{7}{60}-\frac{69 Q^2}{2}\right) (p(p \partial^2 w_4))
    -\frac{321Q^2}{5} (p(\partial p \partial w_4))
    \nonumber\\
    &
    -26 Q^2 (p(\partial^2 p  w_4))
    -\frac{521 Q^2}{30}(\partial p(\partial p  w_4))
    +\left(120 Q^3-\frac{73 Q}{90}\right)(p\partial^3 w_4)
    +\left(147 Q^3-\frac{3 Q}{10}\right)(\partial p\partial^2 w_4)
    \nonumber\\
    &
    +\left(104 Q^3-\frac{Q}{15}\right)(\partial^2 p\partial w_4)
    +\left(\frac{91 Q^3}{3}+\frac{2 Q}{45}\right)(\partial^3 p w_4)
    +\left(-150 Q^4+\frac{3Q^2}{8}+\frac{11}{720}\right)\partial^4 w_4
    \nonumber\\
    &
    +\frac{1}{30}(p(p(w_2(w_2 w_2))))
    -\frac{17 Q}{30}(p(w_2(w_2\partial w_2)))
    -\frac{2 Q}{5}(\partial p(w_2(w_2 w_2)))
    +\frac{1}{5}(p(p(p(p(w_2 w_2)))))
    \nonumber\\
    &
    -\frac{187 Q}{30}(p(p(p(w_2\partial w_2))))
    -\frac{89 Q}{15}(p(p(\partial p(w_2 w_2))))
    +\left(42 Q^2-\frac{1}{20}\right)(p(p(w_2\partial^2 w_2)))
    \nonumber\\
    &+\left(\frac{217   Q^2}{6}-\frac{1}{30}\right)(p(p(\partial w_2\partial w_2)))
    +\left(\frac{1156 Q^2}{15}+\frac{1}{10}\right)(p(\partial p(w_2\partial w_2)))
    +\frac{83 Q^2}{5}(p(\partial^2 p(w_2 w_2)))
    \nonumber\\
    &+\left(\frac{289   Q^2}{10}-\frac{13}{120}\right)(\partial p(\partial p(w_2 w_2)))
    +\left(\frac{Q}{6}-126 Q^3\right)(p(w_2 \partial^3 w_2))
    +\left(\frac{17 Q}{20}-340 Q^3\right)(p(\partial w_2 \partial^2 w_2))
    \nonumber\\
    &+\left(\frac{3 Q}{2}-195Q^3\right)(\partial p(w_2 \partial^2 w_2))
    +\left(-92 Q^3-\frac{7Q}{10}\right)(\partial p(\partial w_2 \partial w_2))
    +\left(20 Q^3-\frac{7Q}{5}\right)(\partial^2 p(w_2\partial w_2))
    \nonumber\\
    &+\left(\frac{191 Q}{360}-\frac{821  Q^3}{30}\right)(\partial^3 p(w_2 w_2))
    +\frac{1}{120}(p(p(p(p(p(p w_2))))))
    -\frac{17Q}{120}(p(p(p(p(p\partial w_2)))))
    \nonumber\\
    &
    -\frac{13Q}{15}(p(p(p(p(\partial pw_2)))))
    +\left(5 Q^2-\frac{1}{10}\right)(p(p(p(p\partial^2 w_2))))
    +\left(\frac{68Q^2}{15}+\frac{1}{20}\right)(p(p(p(\partial p\partial  w_2))))
    \nonumber\\
    &+\frac{23 Q^2}{5}(p(p(p(\partial^2 p  w_2))))
    +\left(\frac{137 Q^2}{6}-\frac{7}{120}\right)(p(p(\partial p(\partial p w_2))))
    +\left(\frac{71Q}{36}-63 Q^3\right)(p(p(p\partial^3w_2)))
    \nonumber\\
    &
    +\left(\frac{63 Q}{20}-\frac{303 Q^3}{2}\right)(p(p(\partial p\partial^2w_2)))
    +\left(28 Q^3-\frac{29 Q}{30}\right)(p(p(\partial^2 p\partial w_2)))
    +\left(\frac{2647Q^3}{30}-\frac{3 Q}{2}\right)(p(\partial p(\partial p\partial w_2)))
    \nonumber
    \\
    &+\left(\frac{19 Q}{40}-\frac{901 Q^3}{30}\right)(p( p(\partial^3 p  w_2)))
    +\left(\frac{322 Q^3}{5}-\frac{8 Q}{15}\right)(p(\partial p(\partial^2 p  w_2)))
    +\left(\frac{161 Q}{60}-\frac{1508 Q^3}{5}\right)(\partial p(\partial p(\partial p w_2)))
    \nonumber\\
    &+\left(300 Q^4-\frac{1267Q^2}{72}+\frac{19}{360}\right)(p(p\partial^4 w_2))
    +\left(348 Q^4-\frac{101 Q^2}{4}-\frac{1}{60}\right)(p(\partial p\partial^3w_2))
    \nonumber\\
    &
    +\left(-286 Q^4-\frac{55Q^2}{6}+\frac{7}{120}\right)(p(\partial^2 p\partial^2w_2))
    +\left(-237 Q^4-\frac{543 Q^2}{20}+\frac{9}{80}\right)(\partial p(\partial p\partial^2w_2))
    \nonumber\\
    &+\left(-\frac{1145Q^4}{3}-\frac{19 Q^2}{180}+\frac{1}{60}\right)(p(\partial^3 p\partial w_2))
    +\left(\frac{1859 Q^2}{60}-4838 Q^4\right)(\partial p(\partial^2 p\partial w_2))
    \nonumber\\
    &
    +\left(-79 Q^4-\frac{287Q^2}{30}+\frac{1}{12}\right)(p(\partial^4pw_2))
    +\left(\frac{58 Q^4}{15}-\frac{593 Q^2}{9}+\frac{131}{360}\right)(\partial p(\partial^3p w_2))
    \nonumber\\
    &
    +\left(-\frac{4157Q^4}{2}-\frac{537 Q^2}{40}+\frac{61}{240}\right)(\partial^2 p(\partial^2p w_2))
    +\left(-1008 Q^5+\frac{118 Q^3}{3}-\frac{13 Q}{400}\right)(p\partial^5 w_2)
    \nonumber\\
    &
    +\left(1410 Q^5+\frac{557 Q^3}{2}-\frac{89 Q}{40}\right)(\partial p \partial^4 w_2)
    +\left(10770 Q^5+\frac{253 Q^3}{12}+\frac{Q}{9}\right)(\partial^2p\partial^3w_2)
    \nonumber\\
    &
    +\left(17338 Q^5+\frac{1417 Q^3}{20}-\frac{379 Q}{240}\right)(\partial^3p\partial^2w_2)
    +\left(12052 Q^5+\frac{997 Q^3}{60}-\frac{7Q}{30}\right)(\partial^4 p \partial w_2)
    \nonumber\\
    &
    +\left(\frac{15042 Q^5}{5}+\frac{22413Q^3}{100}-\frac{1789 Q}{900}\right)(\partial^5pw_2)
    +\widehat{W}_8(p),
\end{align}
}
where $\widehat{W}_8(p)$ is an operator containing a free boson $p$ only:
{\allowdisplaybreaks\small
\begin{align}
\widehat{W}_8(p)
&=
-\frac{17Q^2}{20} (p(p(p(p(p\partial^2 p)))))
+\left(-\frac{Q^2}{8}-\frac{1}{480}\right) (p(p(p(p(\partial p\partial p)))))
\nonumber\\
&
+\left(\frac{799 Q^3}{120}+\frac{119 Q}{1440}\right) (p(p(p(p\partial^3 p))))
+\left(\frac{611 Q^3}{5}-\frac{2 Q}{5}\right) (p(p(p(\partial p\partial^2 p))))
\nonumber\\
&+\left(\frac{71 Q}{120}-\frac{65 Q^3}{2}\right) (p(p(\partial p(\partial p\partial p))))
\nonumber\\
&
+\left(-\frac{239 Q^4}{2}-\frac{89 Q^2}{20}+\frac{1}{24}\right) (p(p(p\partial^4 p)))
+\left(-\frac{13246 Q^4}{15}-\frac{5951 Q^2}{180}+\frac{143}{720}\right) (p(p(\partial p\partial^3 p)))
\nonumber\\
&
+\left(-\frac{6593 Q^4}{4}-\frac{609 Q^2}{80}+\frac{5}{32}\right) (p(p(\partial^2 p\partial^2 p)))
+\left(-\frac{30663 Q^4}{5}+\frac{73 Q^2}{4}+\frac{7}{120}\right) (p(\partial p(\partial p\partial^2 p)))
\nonumber\\
&
+\left(1317 Q^4-\frac{3217 Q^2}{120}+\frac{9}{160}\right) (\partial p(\partial p(\partial p\partial p)))
+\left(\frac{8601 Q^5}{5}+\frac{22073 Q^3}{200}-\frac{1783 Q}{1800}\right) (p(p \partial^5 p))
\nonumber\\
&
+\left(16149 Q^5+\frac{4783 Q^3}{20}-\frac{11 Q}{5}\right) (p(\partial p \partial^4 p))
+\left(31462 Q^5+\frac{1787 Q^3}{20}-\frac{599 Q}{360}\right) (p(\partial^2 p \partial^3 p))
\nonumber\\
&
+\left(\frac{783871 Q^5}{30}+\frac{457841 Q^3}{360}-\frac{7027 Q}{720}\right) (\partial p(\partial p \partial^3 p))
+\left(\frac{174295 Q^5}{2}+\frac{12849 Q^3}{40}-\frac{497 Q}{80}\right) (\partial p(\partial^2 p \partial^2 p))
\nonumber\\
&
+\left(-1428 Q^6-\frac{649 Q^4}{12}+\frac{59 Q^2}{120}\right) (p\partial^6 p)
+\left(-\frac{389346 Q^6}{5}-\frac{543259 Q^4}{100}+\frac{183167 Q^2}{3600}-\frac{137}{7200}\right) (\partial p\partial^5 p)
\nonumber\\
&
+\left(299176 Q^6-\frac{20267 Q^4}{20}+\frac{1097 Q^2}{60}-\frac{1}{20}\right) (\partial^2 p\partial^4 p)
\nonumber\\
&+\left(\frac{1310723 Q^6}{6}-\frac{49307 Q^4}{45}+\frac{21307 Q^2}{864}-\frac{79}{2160}\right) (\partial^3 p\partial^3 p)
\nonumber\\
&
+\left(24771 Q^7+\frac{16054 Q^5}{15}-\frac{4853 Q^3}{280}+\frac{23 Q}{420}\right)\partial^7 p.
\end{align}
}
Other higher-spin currents could also be obtained similarly. The construction of the spin-8 current from the $\W E_7$ algebra is also important to check the consistency of the present method, which is left for future study.

\section{Recursive Construction of \texorpdfstring{$\mathcal{W}BC_r$}{WBCr} Algebra}\label{sec:WBC}

In this section, we study the recursive construction of the $\W BC_r$ algebra for $r\geq 1$. This algebra is generated by spin-$s$ currents with $s=2,4,\dots,2r$. Let $e_1,\dots,e_r$ be an orthonormal basis of ${\mathbb R}^r$. The simple roots and fundamental weights of the Lie algebra $B_r$ are given by
\begin{align}
    \alpha_i(B_r) &= e_i - e_{i+1}, ~~ (i=1,\cdots,r-1),
    \quad
    \alpha_r(B_r)=e_r, \nonumber\\
    \Lambda_i(B_r) &= e_1 + e_2 + \cdots + e_i, ~~(i=1,\cdots,r-1),
    \quad
    \Lambda_r(B_r) = \frac{1}{2}\qty(e_1+e_2+\cdots+e_r).
\end{align}
The Weyl vector and co-Weyl vector are
\begin{equation}
    \rho(B_r) = \frac{2r-1}{2}e_1 + \frac{2r-3}{2}e_2 + \cdots + \frac{1}{2}e_r,
    \quad
    \rho^{\vee}(B_r) = re_1 + (r-1)e_2 + \cdots + e_r.
\end{equation}
The simple roots of $C_r$ are the co-roots $\alpha_i(B_r)^{\vee}$, where the roles of $\rho(B_r)$ and $\rho^{\vee}(B_r)$ are interchanged.

Let $W_s^{(r)}$ denote the spin-$s$ current of the $\W BC_r$ algebra. We introduce $r$ free bosons $\phi=(\phi_1,\dots, \phi_r)$ and their derivatives $p=i\partial \phi$ with $\phi_i=e_i\cdot \phi$. The $\W$-currents should commute with the screening charges $Q_i^{\pm}$ for $i=1,\dots,r$. From the quantum Drinfeld-Sokolov reduction, the spin-2 current $W_2^{(r)}$ is obtained as \eqref{eq:EMtensorWg}, which has the central charge $c(B_r)=r-12(b\rho-\rho^{\vee}/b)^2=r+2r(r+1)(4r-1) -b^2 r(2r-1)(2r+1)-\frac{1}{b^2} 2r(r+1)(2r+1)$. The recurrence structure arises when  one considers the decomposition of the simple roots of $B_r$ into those of $B_{r-1}$ and its orthogonal direction $\te_r$:
\begin{equation}
    \alpha_{1}(B_r) = -\Lambda_1(B_{r-1})+\te_r,
    \quad
    \alpha_{i}(B_r) = \alpha_{i-1}(B_{r-1}) ~~ (i=2,\dots, r).
\end{equation}
We also find that the Weyl vector and the co-Weyl vector are expressed as
\begin{equation}
    \rho(B_r) = \rho(B_{r-1})+\qty(r-\frac{1}{2})\te_r,
    \quad
    \rho^{\vee}(B_r) = \rho^{\vee}(B_{r-1}) + r\te_r
\end{equation}
If we express $W_2^{(r)}$ in terms of $W_2^{(r-1)}$, a free boson $\tp_r$, and their derivatives, the zero commutation relation with the screening charge $Q_{1}^{\pm}$ leads to the recurrence 
relation:
\begin{equation}\label{eq:bcr-rec1}
    W_2^{(r)}
    = W_2^{(r-1)} + \frac{1}{2}(\tp_r\tp_r) - \left( \qty(r-\frac{1}{2})b-\frac{r}{b}\right)\partial \tp_r.
\end{equation}
In the $r=1$ case, we have
\begin{equation}\label{eq:wbc1w2}
    W_2^{(1)} = \frac{1}{2}\qty(\tp_1\tp_1)-\qty(\frac{b}{2}-\frac{1}{b})\partial\tp_1.
\end{equation}
The relation \eqref{eq:bcr-rec1} leads to Eq.~\eqref{eq:EMtensorWg} if we identify 
$\tp_i=p_{r+1-i}$ by reversing the indices of the basis vector $\te_i=e_{r+1-i}$.

Other higher-spin currents of the $\W BC_r$ algebra can be constructed recursively from the $\W BC_{r-1}$ generators and a free boson. Let us illustrate 
this recursive construction for the spin-4 current of the $\W BC_2$ algebra and the spin-4 and 6 currents of the $\W BC_3$ algebra. The latter algebra will be used to construct the $\W F_4$ algebra in the next section.

First, we discuss the spin-4 current of the $\W BC_2$ algebra. This current is built from the $\W BC_1$ current $W_2^{(1)}$ in \eqref{eq:wbc1w2} and a free boson $\widetilde{\phi}_2$. We will denote the former as $w_2$. We define $\tp_2=i\widetilde{\phi}_2$. The most general ansatz for the spin-4 current takes the form:
\begin{align}
    W_4^{(2)} &=
    x_1(w_2w_2)
    + x_2\partial^2w_2
    + x_3(\tp_2(\tp_2w_2))
    + x_4(\partial \tp_2w_2)
    + x_5(\tp_2\partial w_2)
    \nonumber\\
    &\quad
    + x_6(\tp_2(\tp_2(\tp_2\tp_2)))
    + x_7(\tp_2(\tp_2\partial\tp_2))
    + x_8(\tp_2\partial^2\tp_2)
    + x_9(\partial\tp_2\partial\tp_2)
    + x_{10}\partial^3\tp_2.
\end{align}
This ansatz automatically satisfies the zero commutation relation with $Q_2^{\pm}$. The commutation relation with $Q_1^{\pm}$ fixes the coefficients $x_i$ except for $x_1$, $x_2$, and $x_3$. We obtain
\begin{equation}
    W_4^{(2)} = x_1(W_2^{(2)}W_2^{(2)}) + x_2\partial^2W_2^{(2)} + (x_3-x_1)\widehat{W}_4^{(2)},
\end{equation}
where
\begin{align}
    \widehat{W}_4^{(2)} &=
    (\tp_2(\tp_2w_2))+\qty(\frac2b-b)(\partial\tp_2 w_2)+\qty(\frac1b-b)(\tp_2\partial w_2)+\qty(-\frac{1}{b^2}+\frac{3}{2}-\frac{b^2}{2})\qty(\tp_2\partial^2\tp_2)
    \nonumber\\
    &\quad +\qty(\frac{1}{2b^2}-\frac12)(\partial\tp_2\partial\tp_2)+\qty(-\frac{1}{b^3}+\frac{29}{12b}-\frac{23b}{12}+\frac{b^3}{2})\partial^3\tp_2.
\end{align}
$x_3$ fixes the overall normalization of $W_4^{(2)}$, while $x_1$ and $x_2$ parametrize the freedom to add composite fields constructed from the lower-spin generator $W_2^{(2)}$. We can adjust the parameters $x_1$, $x_2$, and $x_3$ so that our $W_4^{(2)}$ matches that of \cite{Ito:1995ny}:
\begin{equation}\label{wbc2w4}
    W_4^{(2)}
    = -4\qty(W_2^{(2)}W_2^{(2)}) + \qty(\frac{5}{b^2}-7+\frac{31b^2}{10})\partial^2W_2^{(2)} + \frac{25}{2}\widehat{W}_4^{(2)}.
\end{equation}

Next, we discuss the spin-4 current $W_4^{(3)}$ of the $\W BC_3$ algebra. This current is built from the $\W BC_2$ currents $W_2^{(2)}$ and $W_4^{(2)}$ in \eqref{wbc2w4}, which are denoted as $w_2$ and $w_4$ below. These currents, together with a free boson $\widetilde{\phi}_3$, are the building blocks of the $\W BC_3$ algebra. We define $\widetilde{p}_3=i\partial\widetilde{\phi}_3$. The most general ansatz for the spin-4 current is written as
\begin{align}
    W_4^{(3)}
    &=
    x_1w_4
    + x_2(w_2w_2)
    + x_3\partial^2w_2
    + x_4(\tp_3(\tp_3w_2))
    + x_5(\partial\tp_3w_2)
    + x_6(\tp_3\partial w_2)
    \nonumber\\
    &\quad
    + x_7(\tp_3(\tp_3(\tp_3\tp_3)))
    + x_8(\tp_3(\tp_3\partial\tp_3))
    + x_9(\tp_3\partial^2\tp_3)
    + x_{10}(\partial\tp_3\partial\tp_3)
    + x_{11}\partial^3\tp_3.
\end{align}
This ansatz automatically satisfies the zero commutation relations with $Q_2^{\pm}$ and $Q_3^{\pm}$. The relation for $Q_1^{\pm}$ determines the coefficients $x_i$ except for the remaining freedom $x_1$, $x_2$, and $x_3$. We obtain
\begin{equation}
    W_4^{(3)} = x_1\widehat{W}_4^{(3)} + x_2(W_2^{(3)}W_2^{(3)}) + x_3\partial^2W_2^{(3)}
\end{equation}
where
\begin{align}
    \widehat{W}_4^{(3)}
    &=
    w_4+\frac{17}{2}(\tp_3(\tp_3w_2))+\qty(\frac{26}{b}-\frac{35b}{2})(\partial\tp_3w_2)+\qty(\frac{25}{2b}-\frac{25b}{2})(\tp_3\partial w_2)-(\tp_3(\tp_3(\tp_3\tp_3)))
    \nonumber\\
    &\quad +\qty(-\frac{12}{b}+10b)(\tp_3(\tp_3\partial \tp_3))+\frac{1}{2}\left(-\frac{65}{b^2}+107-\frac{219b^2}{5}\right)(\tp_3\partial^2\tp_3)
    \nonumber\\
    &\quad +\qty(-\frac{6}{b^2}+\frac{31}{2}-\frac{47b^2}{5})(\partial\tp_3\partial\tp_3)+\qty(-\frac{35}{b^3}+\frac{275}{3b}-\frac{4817b}{60}+\frac{47b^3}{2})\partial^3\tp_3.
\end{align}
Here, $x_2$ and $x_3$ are the coefficients of the terms expressed only by $W_2^{(3)}$, while the overall normalization of $W_4^{(3)}$ is carried by $x_1$. We can adjust the parameters $x_1$, $x_2$, and $x_3$ so that our $W_4^{(3)}$ matches that of \cite{Ito:1995ny}:
\begin{equation}
    W_4^{(3)} = -\frac{16}{25}W_4^{(3)}+\frac{36}{25}(W_2^{(3)}W_2^{(3)})-\frac{6 \left(419 b^4-930 b^2+700\right)}{875 b^2}\partial^2W_2^{(3)}.
\end{equation}

In addition, we discuss the spin-6 current $W_6^{(3)}$ of $\W BC_3$. The most general ansatz for $W_6^{(3)}$ is written with $w_2$, $w_4$, $\tp_3$, and their derivatives as 34 terms. The zero commutation relation with $Q_1^{\pm}$ determines the coefficients of the 34 terms, except for seven parameters. By choosing appropriate free parameters, the result can be written as
\begin{align}
    W_6^{(3)}
    &=
    x_1\widehat{W}_6^{(3)}
    + x_2\partial^2\widehat{W}_4^{(3)}
    + x_3(W_2^{(3)}(W_2^{(3)}W_2^{(3)}))
    + x_4(W_2^{(3)}\partial^2W_2^{(3)})
    \nonumber\\
    &\quad
    + x_5(\partial W_2^{(3)}\partial W_2^{(3)})
    + x_6\partial^4W_2^{(3)}
    + x_7(W_2^{(3)}W_4^{(3)})
\end{align}
where
{\allowdisplaybreaks\small
\begin{align}
    \widehat{W}_6^{(3)}
    &=
    \frac{b}{1-b^2} (\tp_3(\tp_3 w_4)) + \frac{2-b^2}{1-b^2} (\partial{\tp_3} w_4) + (\tp_3 \partial{w_4}) + \frac{4b}{1-b^2} (\tp_3(\tp_3(w_2 w_2))) + \frac{4(2-b^2)}{1-b^2} (\partial{\tp_3}(w_2 w_2)) \nonumber\\
    &\quad + 8 (\tp_3(w_2 \partial{w_2})) - \frac{50 - 70b^2 + 31b^4}{10b(1-b^2)} (\tp_3(\tp_3 \partial^2{w_2})) - \frac{25(1-b^2)}{2b} (\tp_3(\partial{\tp_3} \partial{w_2})) \nonumber\\
    &\quad - \frac{25(2-b^2)}{2b} (\tp_3(\partial^2{\tp_3} w_2)) + \frac{25}{2b} (\partial{\tp_3}(\partial{\tp_3} w_2)) - \frac{25(3-2b^2)(4-3b^2)}{12b^2} (\partial^3{\tp_3} w_2) \nonumber\\
    &\quad - \frac{25(1-b^2)(3-2b^2)}{2b^2} (\partial^2{\tp_3} \partial{w_2}) - \frac{225 - 565b^2 + 507b^4 - 156b^6}{10b^2(1-b^2)} (\partial{\tp_3} \partial^2{w_2}) \nonumber\\
    &\quad - \frac{150 - 190b^2 + 93b^4}{30b^2} (\tp_3 \partial^3{w_2}) + \frac{25(1-b^2)(3-2b^2)(4-3b^2)}{24b^3} (\tp_3 \partial^4{\tp_3}) \nonumber\\
    &\quad - \frac{25(1-b^2)(3-2b^2)}{6b^3} (\partial{\tp_3} \partial^3{\tp_3}) + \frac{25(1-b^2)(2-b^2)}{8b^3} (\partial^2{\tp_3} \partial^2{\tp_3}) \nonumber\\
    &\quad + \frac{5(1-b^2)(4-3b^2)(5-4b^2)(6-5b^2)}{48b^4} \partial^5{\tp_3}.
\end{align}}
For the study of the $\W F_4$ algebra, it is convenient to define the spin-6 current as
\begin{equation}
\label{eq:bc3w6forf4}
    \widetilde{W}_6^{(3)}
    =
    \qty(\frac{3}{2b^3} - \frac{11}{4b} + \frac{5b}{4})\widehat{W}_6^{(3)} + \qty(\frac{1}{2} - \frac{1}{b^2})(W_2^{(3)}W_4^{(3)}).
\end{equation}

We can adjust the parameters $x_i$ so that our $W_6^{(3)}$ matches that of \cite{Ito:1995ny}. The spin-6 current can be alternatively obtained from the OPE coefficient $\{\widehat{W}_4^{(2)}\widehat{W}_4^{(2)}\}_2$, which differs from $\widehat{W}_6^{(3)}$ by a linear combination of the composite fields built from the lower-spin generators and a normalization factor. 

The higher-rank $\W BC_r$ algebras can be constructed similarly. The $\W BC_r$ algebra is recursively constructed from the $\W BC_{r-1}$ generators and a free boson using the zero commutation relation with the screening charge $Q_1^{\pm}$.

We now compute the $\W$-charges of the highest-weight state $|\lambda\rangle$ corresponding to the vertex operator $V=~:e^{i\lambda\cdot \phi}:$ for the generators of the $\W BC_{1,2,3}$ algebra. These charges depend on $\lambda$ only through the shifted momentum $\widetilde{\lambda}=\lambda+b\rho-\rho^{\vee}/b$ and are written in terms of the following Casimir invariants of $B_r$:
\begin{equation}
    C_k = \Tr_{\mathbf{2r+1}}\qty(\widetilde{\lambda}\cdot H)^k, \quad (k=2,4,\ldots,2r),
\end{equation}
where $H$ is the Cartan generator in the $(2r+1)$-dimensional representation of $B_r$. The explicit forms of $H_i$ are given in Appendix~\ref{app:Lie} for $B_1$, $B_2$, and $B_3$, where $C_2$ is normalized as $C_2=2\widetilde{\lambda}^2$.
The $\W$-charge of the spin-2 current $W_2^{(r)}$ is the conformal weight of the vertex operator:
\begin{equation}
    \Delta_2 = \frac{1}{4}C_2+\frac{c(B_r)-r}{24}.
\end{equation}
For $\W BC_2$, the $\W$-charge for the spin-4 current $W_4^{(2)}$ is expressed as
\begin{equation}
    \Delta_4 = \frac{19b^2-120}{160}C_2+\frac{17}{32}C_2^2-\frac{25}{16}C_4+\frac{685b^4-1968b^2+1440}{320}.
\end{equation}
For $\W BC_3$, the $\W$-charges for the spin-4 current $\widehat{W}_4^{(3)}$ and the spin-6 current $\widehat{W}_6^{(3)}$ are expressed as
{\allowdisplaybreaks
\begin{align}
    \Delta_4
    &=
    -\frac{25}{16}C_4+\frac{17}{32}C_2^2-\frac{3\left(351b^4-830b^2+510\right)}{80b^2}C_2
    \nonumber\\
    &\quad +\frac{3\left(21945b^8-104276b^6+186852b^4-149760b^2+45360\right)}{320b^4},
    \\
    \Delta_6
    &=
    \frac{1}{b^2-1}\left(-\frac{25 b}{24}C_6+\frac{25b}{32}C_4C_2-\frac{25 \left(25 b^4-60 b^2+36\right)}{64 b}C_4-\frac{25 b}{192}C_2^3\right.
    \nonumber\\
    &\quad \left.+\frac{25 \left(25 b^4-60 b^2+36\right)}{128 b}C_2^2-\frac{25 \left(625 b^8-3000 b^6+5400 b^4-4320 b^2+1296\right)}{128 b^3}C_2\right.
    \nonumber\\
    &\quad \left.+\frac{25 \left(15625 b^{12}-112500 b^{10}+337500 b^8-540000 b^6+486000 b^4-233280 b^2+46656\right)}{256 b^5}\right).
\end{align}
}
Each $\Delta_s$ is a polynomial in the Casimir invariants whose leading term is $C_s$.

\section{Free Field Realization of \texorpdfstring{$\mathcal{W}F_4$}{WF4} Algebra}\label{sec:WF4}

In this section, we discuss the free field realization of the $\W F_4$ algebra based on the $\W BC_3$ algebra. The $\W F_4$ algebra is generated by the currents $W_s$ of spins $s=2, 6, 8$, and $12$. We now denote the spin-2 current $W_2^{(3)}$, the spin-4 current $\widehat{W}_4^{(3)}$, and the spin-6 current $\widetilde{W}_6^{(3)}$ in \eqref{eq:bc3w6forf4} of the $\W BC_3$ algebra as $w_2$, $w_4$, and $w_6$. These currents and a free boson $\varphi$ are the building blocks of $W_s$. The simple roots of $F_4$ are related to those of $B_3$ as
\begin{equation}
    \alpha_i(F_4) = \alpha_i(B_3), ~~ (i=1,2,3), \quad
    \alpha_4(F_4) = -\Lambda_3(B_3) + \frac{1}{2}e_4.
\end{equation}
Here, $e_4$ is an additional direction orthogonal to the root space of $B_3$, and $\alpha_4$ corresponds to the additional node of the Dynkin diagram. The Weyl vector and co-Weyl vector of $F_4$ are related to those of $B_3$ as
\begin{equation}
    \rho(F_4) = \rho(B_3) + \frac{11}{2}e_4, \quad
    \rho^{\vee}(F_4) = \rho^{\vee}(B_3) + 8e_4.
\end{equation}
The spin-2 current is then obtained as
\begin{equation}
    W_2 = w_2 + \frac{1}{2}(p p) - \qty( \frac{11b}{2} - \frac{8}{b} )\partial p,
\end{equation}
where we define $p=i\partial\varphi$. The central charge is $c(F_4)=-936b^{-2}-468b^2+1324$. $W_2$ satisfies the zero commutation relations with $Q_i^{\pm}~(i=1,2,3,4)$ in \eqref{eq:scrQ}.

Let us construct the spin-6 current $W_6$. The most general ansatz for $W_6$ is written as
\begin{align}
    W_6
    &=
    x_2w_6+x_3(p(pw_4))+x_4(\partial pw_4)+x_5(p\partial w_4)+x_6\partial^2w_4+x_7(w_2(w_2w_2))
    \nonumber\\
    &\quad +x_8(w_2\partial^2w_2)+x_9(\partial w_2\partial w_2)+x_{10}(\partial p(w_2w_2))+x_{11}(p(w_2\partial w_2))+x_{12}\partial^4w_2
    \nonumber\\
    &\quad +x_{13}(p(p(p(pw_2))))+x_{14}(p(p(p\partial w_2)))+x_{15}(p(p(\partial pw_2)))+x_{16}(p(p\partial^2w_2))
    \nonumber\\
    &\quad +x_{17}(p(\partial p\partial w_2))+x_{18}(p(\partial^2pw_2))+x_{19}(\partial p(\partial pw_2))+x_{20}(\partial^3pw_2)+x_{21}(\partial^2p\partial w_2)
    \nonumber\\
    &\quad +x_{22}(\partial p\partial^2w_2)+x_{23}(p\partial^3w_2)+x_{24}(w_2w_4)+x_{25}(p(p(w_2w_2)))
    \nonumber\\
    &\quad +x_{26}(p(p(p(p(pp)))))+x_{27}(p(p(p(p\partial p))))+x_{28}(p(p(p\partial^2p)))+x_{29}(p(p(\partial p\partial p)))
    \nonumber\\
    &\quad +x_{30}(p(p\partial^3p))+x_{31}(p(\partial p\partial^2p))+x_{32}(\partial p(\partial p\partial p))+x_{33}(p\partial^4p)+x_{34}(\partial p\partial^3p)
    \nonumber\\
    &\quad +x_{35}(\partial^2p\partial^2p)+x_{36}\partial^5p.
\end{align}
This current satisfies the zero commutation relations with $Q_i^{\pm}~(i=1,2,3)$ automatically. Imposing the relations for $Q_4^{\pm}$ fixes the coefficients $x_i$ except for $x_2,x_7,x_8,x_9$, and $x_{12}$. We then obtain
\begin{equation}
    W_6
    =
    x_2\qty(\frac{1}{24}\widehat{W}_6)+x_7(W_2(W_2W_2))+x_8(W_2\partial^2W_2)+x_9(\partial W_2\partial W_2)+x_{12}\partial^4W_2
\end{equation}
where
{\allowdisplaybreaks{\small
\begin{align}
    \widehat{W}_6
    &=
    24w_6+5\left(\frac{6}{b^2}-5\right)(p(pw_4))+\qty(\frac{48}{b^3}-\frac{82}{b}+35b)(\partial pw_4)+6\qty(\frac{18}{b^3}-\frac{27}{b}+10b)(p\partial w_4)
    \nonumber\\
    &\quad +\qty(\frac{108}{b^4}-\frac{240}{b^2}+179-45b^2)\partial^2w_4+\qty(\frac{444}{b^3}-\frac{580}{b}+175b)(\partial p(w_2w_2))
    \nonumber\\
    &\quad +\qty(\frac{114}{b^3}-\frac{221}{b}+105b)(p(w_2\partial w_2))-13\left(\frac{3}{b^2}-\frac{5}{2}\right)(p(p(p(pw_2))))
    \nonumber\\
    &\quad -25\left(\frac{15}{b^3}-\frac{43}{2b}+\frac{15b}{2}\right)(p(p(p\partial w_2)))-\qty(\frac{498}{b^3}-\frac{823}{b}+340b)(p(p(\partial pw_2)))
    \nonumber\\
    &\quad -\qty(\frac{1062}{b^4}-\frac{2112}{b^2}+\frac{13501}{10}-273b^2)(p(p\partial^2w_2))
    -\qty(\frac{3228}{b^4}-\frac{7115}{b^2}+\frac{10453}{2}-\frac{2565b^2}{2})(p(\partial p\partial w_2))
    \nonumber\\
    &\quad -\qty(\frac{1824}{b^4}-\frac{4001}{b^2}+\frac{29117}{10}-\frac{1407b^2}{2})(p(\partial^2pw_2))+\qty(\frac{1317}{b^4}-\frac{4121}{2b^2}+\frac{8643}{10}-\frac{103b^2}{2})(\partial p(\partial pw_2))
    \nonumber\\
    &\quad +\qty(-\frac{2592}{b^5}+\frac{15471}{2b^3}-\frac{172267}{20b}+\frac{254191b}{60}-\frac{1551b^3}{2})(\partial^3pw_2)
    \nonumber\\
    &\quad +\qty(-\frac{4824}{b^5}+\frac{15324}{b^3}-\frac{18207}{b}+\frac{19199b}{2}-\frac{3795b^3}{2})(\partial^2p\partial w_2)
    \nonumber\\
    &\quad +\qty(-\frac{4032}{b^5}+\frac{11310}{b^3}-\frac{59417}{5b}+\frac{27589b}{5}-\frac{1893b^3}{2})(\partial p\partial^2w_2)
    \nonumber\\
    &\quad +\qty(-\frac{990}{b^5}+\frac{2244}{b^3}-\frac{16733}{10b}+\frac{2036b}{5}+\frac{3b^3}{2})(p\partial^3w_2)+2\qty(\frac{6}{b^2}-1)(w_2w_4)
    \nonumber\\
    &\quad +7\left(\frac{6}{b^2}-5\right)(p(p(w_2w_2)))+\qty(\frac{6}{b^2}-5)(p(p(p(p(pp)))))+3\qty(\frac{96}{b^3}-\frac{146}{b}+55b)(p(p(p(p\partial p))))
    \nonumber\\
    &\quad +\qty(\frac{888}{b^4}-\frac{4099}{2b^2}+\frac{31383}{20}-\frac{1593b^2}{4})(p(p(p\partial^2p)))+\qty(\frac{5889}{2b^4}-\frac{24941}{4b^2}+\frac{87253}{20}-\frac{2019b^2}{2})(p(p(\partial p\partial p)))
    \nonumber\\
    &\quad +\qty(\frac{1854}{b^5}-\frac{22029}{4b^3}+\frac{244233}{40b}-\frac{119853b}{40}+\frac{2199b^3}{4})(p(p\partial^3p))
    \nonumber\\
    &\quad +\qty(\frac{4884}{b^5}-\frac{13286}{b^3}+\frac{134439}{10b}-\frac{119663b}{20}+\frac{3933b^3}{4})(p(\partial p\partial^2p))
    \nonumber\\
    &\quad +\qty(\frac{7284}{b^5}-\frac{38183}{2b^3}+\frac{368423}{20b}-\frac{154773b}{20}+1192b^3)(\partial p(\partial p\partial p))
    \nonumber\\
    &\quad +\qty(\frac{3690}{b^6}-\frac{26199}{2b^4}+\frac{184543}{10b^2}-\frac{515347}{40}+\frac{178081b^2}{40}-\frac{2433b^4}{4})(p\partial^4p)
    \nonumber\\
    &\quad +\qty(-\frac{2826}{b^6}+\frac{13244}{b^4}-\frac{1413061}{60b^2}+\frac{2419811}{120}-\frac{335483b^2}{40}+\frac{2727b^4}{2})(\partial p\partial^3p)
    \nonumber\\
    &\quad +\qty(-\frac{2601}{b^6}+\frac{16473}{b^4}-\frac{688737}{20b^2}+\frac{328609}{10}-\frac{297281b^2}{20}+\frac{5193b^4}{2})(\partial^2p\partial^2p)
    \nonumber\\
    &\quad +\qty(\frac{4320}{b^7}-\frac{14826}{b^5}+\frac{96447}{5b^3}-\frac{3289751}{300b}+\frac{967907b}{600}+\frac{33543b^3}{40}-264b^5)\partial^5p.
\end{align}
}}
We obtain the higher-spin generators $W_8$ and $W_{12}$ as
\begin{equation}
\label{eq:WF4W8W12}
    W_8 = \{\widetilde{W}_6\widetilde{W}_6\}_4,
    \quad
    W_{12}=\{\widetilde{W}_6W_8\}_2,
\end{equation}
where we define $\widetilde{W}_6=\frac{b^2}{6(6-5b^2)}\widehat{W}_6$.

Let us calculate the $\W$-charges of the $\W F_4$ generators. We define $\widetilde{\lambda}=\lambda+b\rho-\rho^{\vee}/b$. The Casimir invariants are defined as $C_k=\Tr_{\mathbf{26}}\qty(\widetilde{\lambda}\cdot H)^k~(k=2,4,6,8,10,12)$, where $H$ is the Cartan generator in the 26-dimensional representation of $F_4$, whose explicit form is given in Appendix~\ref{app:Lie}. The $\W$-charges associated with the spin-2 and spin-6 generators $W_2$, $\widetilde{W}_6$ are obtained as
\begin{align}
    \Delta_2
    &=
    \frac{C_2}{12}-\frac{39b^2}{2}-\frac{39}{b^2}+55=\frac{C_2}{12}+\frac{c(F_4)-4}{24},
    \\
    \Delta_6
    &=
    \frac{25}{72}C_6-\frac{59}{20736}C_2^3-\frac{5129b^4-14540b^2+10220}{17280b^2}C_2^2
    \nonumber\\
    &\quad +\frac{41907b^8-265792b^6+620294b^4-633640b^2+239610}{480b^4}C_2
    \nonumber\\
    &\quad -\frac{\left(73b^4-206b^2+146\right)\left(12571b^8-105490b^6+292302b^4-335800b^2+138320\right)}{160b^6}.
\end{align}
The $\W$-charges for the spin-8 and spin-12 generators in \eqref{eq:WF4W8W12} are given in Appendix~\ref{app:WF4Charge}. Each $\Delta_k$ is a polynomial in the Casimir invariants whose leading term is $C_k$.

\section{Conclusion and Discussion}\label{sec:discussion}

In this paper, we have studied the free field realizations of the $\W$-algebras associated with the exceptional Lie algebras $E_6$, $E_7$, $E_8$, and $F_4$. Our construction is based on a recursive structure in which a $\W$-algebra is obtained from a $\W$-algebra of one lower rank together with an additional free boson. We constructed the $\W E_6$ algebra from the $\W D_5$ algebra and a free boson, and showed that the algebra is equivalent to the one obtained from the $\W A_5$ algebra, up to a change of basis of the free fields. We have also realized the $\W E_7$ algebra from the $\W D_6$ and $\W E_6$ algebras, and shown that these two realizations yield equivalent results. For the $\W E_8$ algebra, the spin-2 and spin-8 generators from the $\W D_7$ algebra are constructed. In addition, the recursive construction of the $\W BC_r$ algebra is studied, and the $\W F_4$ algebra is constructed based on the $\W BC_3$ algebra. The $\W$-charges of the generators in the $\W E_6$, $\W E_7$, $\W BC_2$, $\W BC_3$, and $\W F_4$ algebras have also been derived.

Our work provides an efficient computational framework for studying the $\W$-algebras associated with the exceptional Lie algebras. An important advantage of the free field realization is that it provides an explicit description of the generators in terms of free bosons. This makes it possible to perform explicit calculations of OPEs and $\W$-charges. Also, the free field realization provides a starting point for constructing representations of exceptional $\W$-algebras. The expressions for the generators and their eigenvalues on vertex operators obtained in this paper may be useful for investigating the structure of highest-weight representations and their associated representation spaces.

The representation theory of $\W$-algebras is also important in the study of correlation functions in two-dimensional conformal field theories. In the free field realizations, screening charges provide a useful framework for constructing correlation functions. It would be interesting to investigate how the screening charges and recursive structures developed in our work can be used to study conformal blocks and correlation functions associated with exceptional $\W$-algebras. Such a formulation may provide a concrete approach to the representation theory of exceptional $\W$-algebras, in which explicit computational tools are comparatively limited at present.

Another important application is the AGT correspondence \cite{Keller:2011ek}, which relates two-dimensional conformal field theories with $\W$-algebra symmetry to supersymmetric gauge theories. For the classical Lie algebras, the corresponding $\W$-algebras play an important role in the description of conformal blocks and their relation to instanton partition functions of supersymmetric gauge theories. Such correspondences have not been extended to exceptional gauge groups so far. The explicit free field realizations obtained in this work may provide a useful framework for investigating conformal field theories with exceptional $\W$-algebra symmetry and their relations to gauge theories associated with exceptional Lie groups \cite{Drinfeld:1984qv}.

The exceptional $\W$-algebra can serve as a tool to investigate integrable systems with exceptional Lie algebras. The existence of higher-spin currents in $\W$-algebras implies the existence of infinite families of commuting integrals of motion in integrable deformations of conformal field theories. The explicit free field realizations constructed in this paper provide a practical framework for investigating such integrable structures for the exceptional Lie algebras. In particular, explicit expressions for the $\W$-currents may be useful for constructing and analyzing local and nonlocal integrals of motion and for studying their spectra and eigenvalues. It would be interesting to study how the recursive structures found in the present work are reflected in the corresponding integrable systems and their conserved quantities.

The ODE/IM correspondence provides important motivation for the present work \cite{Dorey:2000ma,Dorey:2006an,Sun:2012xw,Ito:2015nla,Masoero:2015lga,Ito:2020htm,Ide:2026cxp}. For exceptional Lie algebras, the correspondence between the integrals of motion in conformal field theories with $\W$-algebra symmetry and the WKB periods of differential equations associated with affine Lie algebras is highly non-trivial. The explicit construction of exceptional $\W$-currents developed in this paper provides a concrete framework for studying the integrals of motion in the conformal field theories. In particular, it would be interesting to study whether the recursive structures of exceptional $\W$-algebras are reflected in the corresponding quantum spectral problems and their WKB analysis. The free field realization may also facilitate explicit computations of the integrals of motion and their eigenvalues, which are expected to correspond to the period integrals obtained from the differential equations.

Several problems remain for future investigation. A complete construction of all generators of the $\W E_8$ algebra is an open problem. It would also be interesting to clarify the representation theory for the exceptional $\W$-algebras in this work and to investigate their conformal blocks and correlation functions. Further applications to the AGT correspondence and the ODE/IM correspondence may reveal new structures that are unique to exceptional Lie algebras. We hope that the recursive construction and free field methods developed in this work will provide useful tools for these investigations.

\section*{Acknowledgments}

We would like to thank Naozumi Tanabe, Wataru Kono for their useful discussions and comments. D.I. is supported by the Tsubame Scholarship for Doctoral Students at Institute of Science Tokyo.

\appendix

\section{Properties of Lie Algebras}\label{app:Lie}

\begin{figure}[h]
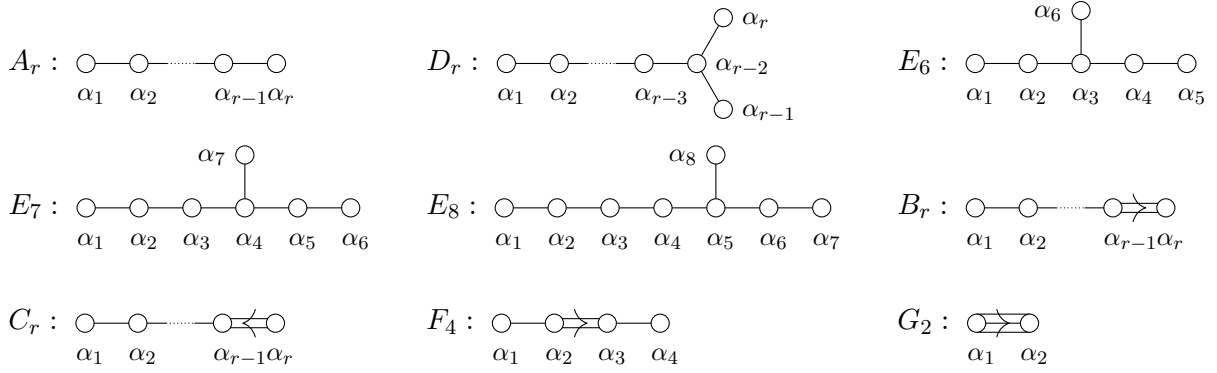

    \centering
    \begin{tabular}{@{}l@{\qquad}l@{\qquad}l@{}}
    $A_r:$
    \dynkin[
        labels = {1,2,r-1,r},
        ordering = Kac,
        text style/.style = {scale=0.9},
        label macro/.code = {\ensuremath{\alpha_{\drlap{#1}}}}
    ]{A}{}
    &
    $D_r:$
    \dynkin[
        labels = {1,2,r-3,r-2,r-1,r},
        label directions = {,,,right,right,right},
        ordering = Kac,
        text style/.style = {scale=0.9},
        label macro/.code = {\ensuremath{\alpha_{\drlap{#1}}}}
    ]{D}{oo.oooo}
    &
    $E_6:$
    \dynkin[
        label,
        label directions = {,,,,,left},
        ordering = Kac,
        text style/.style = {scale=0.9},
        label macro/.code = {\ensuremath{\alpha_{\drlap{#1}}}}
    ]{E}{6}
    \\[0.7cm]
    $E_7:$
    \dynkin[
        labels = {6,5,4,3,2,1,7},
        ordering = Kac,backwards,
        text style/.style = {scale=0.9},
        label macro/.code = {\ensuremath{\alpha_{\drlap{#1}}}}
    ]{E}{7}
    &
    $E_8:$
    \dynkin[
        label,
        ordering = Kac,backwards,
        text style/.style = {scale=0.9},
        label macro/.code = {\ensuremath{\alpha_{\drlap{#1}}}}
    ]{E}{8}
    &
    $B_r:$
    \dynkin[
        labels = {1,2,r-1,r},
        text style/.style = {scale=0.9},
        label macro/.code={\alpha_{\drlap{#1}}}
    ]{B}{oo.oo}
    \\[1.0cm]
    $C_r:$
    \dynkin[
        labels = {1,2,r-1,r},
        text style/.style = {scale=0.9},
        label macro/.code={\alpha_{\drlap{#1}}}
    ]{C}{oo.oo}
    &
    $F_4:$
    \dynkin[
        label,
        text style/.style = {scale=0.9},
        label macro/.code={\alpha_{\drlap{#1}}}
    ]{F}{4}
    &
    $G_2:$
    \dynkin[
        labels={2,1},backwards,
        text style/.style = {scale=0.9},
        label macro/.code={\alpha_{\drlap{#1}}}
    ]{G}{2}
    \end{tabular}
    \caption{Dynkin diagrams of the simple Lie algebras with our labeling conventions for the simple roots.}
    \label{fig:dynkin}
\end{figure}

In this appendix, we summarize the basic data of the Lie algebras studied in this paper. In Table~\ref{tab:data}, we list the degrees of the independent Casimir invariants, which coincide with the spins of the generators of the corresponding $\W$-algebra. In Figure~\ref{fig:dynkin}, we show the Dynkin diagrams and the labeling of the simple roots.

$H_i \coloneq \alpha_i\cdot H~(i=1,\dots,r)$ denote the generators of the Cartan subalgebra. We use the matrix representation given in the Appendix of \cite{Ito:2020htm}.

\begin{table}[h]
\renewcommand{\arraystretch}{1.3}
\centering
\begin{tabular}{|c|l||c|l|}
\hline
    $\g$ & degrees & $\g$ & degrees \\
    \hline
    $A_r$ & $2,3,\ldots,r+1$ & $E_6$ & $2,5,6,8,9,12$ \\
    $B_r$ & $2,4,\dots,2r$ & $E_7$ & $2,6,8,10,12,14,18$ \\
    $C_r$ & $2,4,\dots,2r$ & $E_8$ & $2,8,12,14,18,20,24,30$ \\
    $D_r$ & $2,4,\ldots,2r-2; r$~~~ & $F_4$ & $2,6,8,12$ \\
    & & $G_2$ & $2,6$ \\
    \hline
\end{tabular}
\caption{The degrees of the Casimir invariants.}
\label{tab:data}
\end{table}

\newpage

\subsection{\texorpdfstring{$E_6$}{E6}}

In the 27-dimensional representation, the Cartan generators are
{\small
\begin{align}
H_1&=E_{1, 1} - E_{2, 2} + E_{12, 12} + E_{14, 14} - E_{15, 15} + 
 E_{16, 16} - E_{17, 17} + E_{18, 18} - E_{19, 19} + E_{20, 20} - 
 E_{21, 21} - E_{22, 22},
 \nonumber\\
 H_2&=E_{2, 2} - E_{3, 3} + E_{10, 10} + E_{11, 11} - E_{12, 12} + 
 E_{13, 13} - E_{14, 14} - E_{16, 16} + E_{21, 21} + E_{22, 22} - 
 E_{23, 23} - E_{24, 24},
 \nonumber\\
 H_3&=E_{3, 3} - E_{4, 4} + E_{8, 8} + E_{9, 9} - E_{10, 10} - E_{11, 11} + 
 E_{16, 16} - E_{18, 18} + E_{19, 19} - E_{21, 21} + E_{24, 24} - 
 E_{25, 25},
 \nonumber\\
 H_4&=E_{4, 4} - E_{5, 5} + E_{6, 6} - E_{8, 8} + E_{11, 11} - E_{13, 13} + 
 E_{14, 14} - E_{16, 16} + E_{17, 17} - E_{19, 19} + E_{25, 25} - 
 E_{26, 26},
 \nonumber\\
 H_5&=E_{5, 5} - E_{7, 7} + E_{8, 8} - E_{9, 9} + E_{10, 10} - E_{11, 11} + 
 E_{12, 12} - E_{14, 14} + E_{15, 15} - E_{17, 17} + E_{26, 26} - 
 E_{27, 27},
 \nonumber\\
 H_6&=E_{4, 4} + E_{5, 5} - E_{6, 6} + E_{7, 7} - E_{8, 8} - E_{9, 9} + 
 E_{18, 18} - E_{20, 20} + E_{21, 21} - E_{22, 22} + E_{23, 23} - 
 E_{24, 24}.
\end{align}
}

\subsection{\texorpdfstring{$E_7$}{E7}}

In the 56-dimensional representation, the Cartan generators are
{\allowdisplaybreaks\small
\begin{align}
H_1&=E_{1, 1} - E_{2, 2} + E_{16, 16} + E_{18, 18} - E_{19, 19} + 
 E_{21, 21} - E_{22, 22} + E_{23, 23} - E_{25, 25} + E_{26, 26} + 
 E_{27, 27} - E_{28, 28}
\nonumber\\ 
&
 + E_{29, 29} - E_{30, 30} - E_{31, 31} + 
 E_{32, 32} - E_{33, 33} + E_{35, 35} - E_{36, 36} + E_{38, 38} - 
 E_{39, 39} - E_{42, 42} + E_{55, 55} - E_{56, 56},
 \nonumber\\
 H_2&=E_{2, 2} - E_{3, 3} + E_{13, 13} + E_{15, 15} - E_{16, 16} + 
 E_{17, 17} - E_{18, 18} + E_{20, 20} - E_{21, 21} - E_{23, 23} + 
 E_{24, 24} - E_{27, 27}
\nonumber\\ 
&
 + E_{31, 31} + E_{33, 33} - E_{34, 34} + 
 E_{36, 36} - E_{37, 37} + E_{39, 39} - E_{40, 40} + E_{42, 42} - 
 E_{43, 43} - E_{45, 45} + E_{54, 54} - E_{55, 55},
 \nonumber\\
 H_3&=E_{3, 3} - E_{4, 4} + E_{11, 11} + E_{12, 12} - E_{13, 13} + 
 E_{14, 14} - E_{15, 15} - E_{17, 17} + E_{23, 23} - E_{26, 26} + 
 E_{27, 27} + E_{28, 28} 
\nonumber\\ 
&
- E_{29, 29} + E_{30, 30} - E_{31, 31} - 
 E_{33, 33} + E_{40, 40} - E_{41, 41} + E_{43, 43} - E_{44, 44} + 
 E_{45, 45} - E_{47, 47} + E_{53, 53} - E_{54, 54},
 \nonumber\\
   H_4&=E_{4, 4} - E_{5, 5} + E_{9, 9} + E_{10, 10} - E_{11, 11} - 
 E_{12, 12} + E_{17, 17} - E_{20, 20} + E_{21, 21} - E_{23, 23} + 
 E_{25, 25} - E_{28, 28} 
\nonumber\\ 
&
+ E_{29, 29} - E_{32, 32} + E_{33, 33} - 
 E_{36, 36} + E_{37, 37} - E_{40, 40} + E_{44, 44} - E_{46, 46} + 
 E_{47, 47} - E_{49, 49} + E_{52, 52} - E_{53, 53},
 \nonumber\\
H_5&=E_{5, 5} - E_{6, 6} + E_{7, 7} - E_{9, 9} + E_{12, 12} - E_{14, 14} + 
 E_{15, 15} - E_{17, 17} + E_{18, 18} - E_{21, 21} + E_{22, 22} - 
 E_{25, 25} + E_{32, 32} 
\nonumber\\ 
&
- E_{35, 35} + E_{36, 36} - E_{39, 39} + 
 E_{40, 40} + E_{41, 41} - E_{43, 43} - E_{44, 44} + E_{49, 49} + 
 E_{50, 50} - E_{51, 51} - E_{52, 52},
 \nonumber\\
 H_6&=E_{6, 6} - E_{8, 8} + E_{9, 9} - E_{10, 10} + E_{11, 11} - 
 E_{12, 12} + E_{13, 13} - E_{15, 15} + E_{16, 16} - E_{18, 18} + 
 E_{19, 19} - E_{22, 22} 
\nonumber\\ 
&
+ E_{35, 35} - E_{38, 38} + E_{39, 39} - 
 E_{42, 42} + E_{43, 43} + E_{44, 44} - E_{45, 45} + E_{46, 46} - 
 E_{47, 47} + E_{48, 48} - E_{49, 49} - E_{50, 50},
 \nonumber\\
 H_7&=E_{5, 5} + E_{6, 6} - E_{7, 7} + E_{8, 8} - E_{9, 9} - E_{10, 10} + 
 E_{20, 20} + E_{23, 23} - E_{24, 24} + E_{26, 26} - E_{27, 27} + 
 E_{28, 28} - E_{29, 29} 
\nonumber\\ 
&
- E_{30, 30} + E_{31, 31} - E_{33, 33} + 
 E_{34, 34} - E_{37, 37} + E_{46, 46} - E_{48, 48} + E_{49, 49} - 
 E_{50, 50} + E_{51, 51} - E_{52, 52}.
\end{align}
}

\subsection{\texorpdfstring{$E_8$}{E8}}

In the 248-dimensional representation, the Cartan generators are given by
{\small
{\allowdisplaybreaks
\begin{align}
 H_1&=E_{1, 1} - E_{2, 2} + E_{18, 18} - E_{21, 21} + E_{22, 22} - 
 E_{25, 25} + E_{26, 26} - E_{29, 29} + E_{30, 30} - E_{33, 33} + 
 E_{34, 34} + E_{35, 35} 
 \nonumber\\ &
- E_{37, 37} - E_{38, 38} + E_{39, 39} + 
 E_{40, 40} - E_{42, 42} - E_{43, 43} + E_{44, 44} + E_{45, 45} - 
 E_{48, 48} - E_{49, 49} + E_{50, 50} + E_{51, 51} 
 \nonumber\\ &- E_{54, 54} - 
 E_{55, 55} + E_{56, 56} + E_{57, 57} + E_{58, 58} - E_{60, 60} - 
 E_{61, 61} - E_{62, 62} + E_{63, 63} + E_{64, 64} - E_{67, 67} - 
 E_{68, 68} 
 \nonumber\\ &
+ E_{69, 69} + E_{70, 70} - E_{74, 74} - E_{75, 75} + 
 E_{76, 76} + E_{77, 77} - E_{81, 81} - E_{82, 82} + E_{83, 83} + 
 E_{84, 84} - E_{89, 89} - E_{90, 90} 
 \nonumber\\ &
+ E_{91, 91} - E_{97, 97} + 
 E_{98, 98} - E_{104, 104} + E_{105, 105} - E_{111, 111} + 
 E_{112, 112} - E_{119, 119} + 2 E_{120, 120} - 2 E_{129, 129}  \nonumber\\ &+ 
 E_{130, 130} - E_{137, 137} + E_{138, 138} - E_{144, 144} + 
 E_{145, 145} - E_{151, 151} + E_{152, 152} - E_{158, 158} + 
 E_{159, 159} + E_{160, 160}  \nonumber\\ &- E_{165, 165} - E_{166, 166} + 
 E_{167, 167} + E_{168, 168} - E_{172, 172} - E_{173, 173} + 
 E_{174, 174} + E_{175, 175} - E_{179, 179} - E_{180, 180}  \nonumber\\ &+ 
 E_{181, 181} + E_{182, 182} - E_{185, 185} - E_{186, 186} + 
 E_{187, 187} + E_{188, 188} + E_{189, 189} - E_{191, 191} - 
 E_{192, 192} - E_{193, 193}  \nonumber\\ &+ E_{194, 194} + E_{195, 195} - 
 E_{198, 198} - E_{199, 199} + E_{200, 200} + E_{201, 201} - 
 E_{204, 204} - E_{205, 205} + E_{206, 206} + E_{207, 207}  \nonumber\\ &- 
 E_{209, 209} - E_{210, 210} + E_{211, 211} + E_{212, 212} - 
 E_{214, 214} - E_{215, 215} + E_{216, 216} - E_{219, 219} + 
 E_{220, 220} - E_{223, 223}  \nonumber\\ &+ E_{224, 224} - E_{227, 227} + 
 E_{228, 228} - E_{231, 231} + E_{247, 247} - E_{248, 248},
\end{align}}
\begin{align}
 H_2&=E_{2, 2} - E_{3, 3} + E_{15, 15} - E_{18, 18} + E_{19, 19} - 
 E_{22, 22} + E_{23, 23} - E_{26, 26} + E_{27, 27} - E_{30, 30} + 
 E_{31, 31} + E_{32, 32} \nonumber\\ & - E_{34, 34} - E_{35, 35} + E_{36, 36} - 
 E_{40, 40} + E_{41, 41} + E_{42, 42} - E_{45, 45} + E_{46, 46} - 
 E_{47, 47} + E_{48, 48} - E_{51, 51} + E_{52, 52} \nonumber\\ &- E_{53, 53} + 
 E_{54, 54} - E_{58, 58} - E_{59, 59} + E_{60, 60} + E_{61, 61} - 
 E_{65, 65} - E_{66, 66} + E_{67, 67} + E_{68, 68} - E_{72, 72} - 
 E_{73, 73} \nonumber\\ &+ E_{74, 74} + E_{75, 75} - E_{79, 79} - E_{80, 80} + 
 E_{81, 81} + E_{82, 82} - E_{87, 87} - E_{88, 88} + E_{89, 89} + 
 E_{90, 90} - E_{95, 95} - E_{96, 96} \nonumber\\ &+ E_{97, 97} - E_{103, 103} + 
 E_{104, 104} - E_{110, 110} + E_{111, 111} + E_{112, 112} - 
 E_{118, 118} + 2 E_{119, 119} - E_{120, 120} + E_{129, 129} \nonumber\\ &- 
 2 E_{130, 130} + E_{131, 131} - E_{137, 137} - E_{138, 138} + 
 E_{139, 139} - E_{145, 145} + E_{146, 146} - E_{152, 152} + 
 E_{153, 153} + E_{154, 154} \nonumber\\ &- E_{159, 159} - E_{160, 160} + 
 E_{161, 161} + E_{162, 162} - E_{167, 167} - E_{168, 168} + 
 E_{169, 169} + E_{170, 170} - E_{174, 174} - E_{175, 175} \nonumber\\ &+ 
 E_{176, 176} + E_{177, 177} - E_{181, 181} - E_{182, 182} + 
 E_{183, 183} + E_{184, 184} - E_{188, 188} - E_{189, 189} + 
 E_{190, 190} + E_{191, 191} \nonumber\\ &- E_{195, 195} + E_{196, 196} - 
 E_{197, 197} + E_{198, 198} - E_{201, 201} + E_{202, 202} - 
 E_{203, 203} + E_{204, 204} - E_{207, 207} - E_{208, 208}\nonumber\\ & + 
 E_{209, 209} - E_{213, 213} + E_{214, 214} + E_{215, 215} - 
 E_{217, 217} - E_{218, 218} + E_{219, 219} - E_{222, 222} + 
 E_{223, 223} - E_{226, 226} \nonumber\\ &+ E_{227, 227} - E_{230, 230} + 
 E_{231, 231} - E_{234, 234} + E_{246, 246} - E_{247, 247},
\end{align}
{\allowdisplaybreaks
\begin{align}
 H_3&=E_{3, 3} - E_{4, 4} + E_{13, 13} - E_{15, 15} + E_{16, 16} - 
 E_{19, 19} + E_{20, 20} - E_{23, 23} + E_{24, 24} - E_{27, 27} + 
 E_{28, 28} - E_{32, 32}  \nonumber\\ &+ E_{34, 34} + E_{37, 37} - E_{39, 39} + 
 E_{40, 40} - E_{42, 42} + E_{43, 43} - E_{44, 44} + E_{45, 45} - 
 E_{48, 48} + E_{49, 49} - E_{50, 50} + E_{51, 51}  \nonumber\\ &- E_{54, 54} + 
 E_{55, 55} - E_{57, 57} + E_{58, 58} - E_{61, 61} + E_{62, 62} - 
 E_{64, 64} + E_{65, 65} - E_{68, 68} - E_{71, 71} + E_{72, 72} - 
 E_{78, 78}  \nonumber\\ &+ E_{79, 79} + E_{80, 80} - E_{85, 85} - E_{86, 86} + 
 E_{87, 87} + E_{88, 88} - E_{93, 93} - E_{94, 94} + E_{95, 95} + 
 E_{96, 96} - E_{101, 101} - E_{102, 102}  \nonumber\\ &+ E_{103, 103} + 
 E_{105, 105} - E_{109, 109} + E_{110, 110} + E_{111, 111} - 
 E_{112, 112} - E_{117, 117} + 2 E_{118, 118} - E_{119, 119} + 
 E_{130, 130}  \nonumber\\ &- 2 E_{131, 131} + E_{132, 132} + E_{137, 137} - 
 E_{138, 138} - E_{139, 139} + E_{140, 140} - E_{144, 144} - 
 E_{146, 146} + E_{147, 147} + E_{148, 148}  \nonumber\\ &- E_{153, 153} - 
 E_{154, 154} + E_{155, 155} + E_{156, 156} - E_{161, 161} - 
 E_{162, 162} + E_{163, 163} + E_{164, 164} - E_{169, 169} - 
 E_{170, 170}  \nonumber\\ &+ E_{171, 171} - E_{177, 177} + E_{178, 178} + 
 E_{181, 181} - E_{184, 184} + E_{185, 185} - E_{187, 187} + 
 E_{188, 188} - E_{191, 191} + E_{192, 192}  \nonumber\\ &- E_{194, 194} + 
 E_{195, 195} - E_{198, 198} + E_{199, 199} - E_{200, 200} + 
 E_{201, 201} - E_{204, 204} + E_{205, 205} - E_{206, 206} + 
 E_{207, 207}  \nonumber\\ &- E_{209, 209} + E_{210, 210} - E_{212, 212} - 
 E_{215, 215} + E_{217, 217} - E_{221, 221} + E_{222, 222} - 
 E_{225, 225} + E_{226, 226} - E_{229, 229} \nonumber\\ & + E_{230, 230} - 
 E_{233, 233} + E_{234, 234} - E_{236, 236} + E_{245, 245} - 
 E_{246, 246},
\end{align}}
\begin{align} 
 H_4&=E_{4, 4} - E_{5, 5} + E_{11, 11} - E_{13, 13} + E_{14, 14} - 
 E_{16, 16} + E_{17, 17} - E_{20, 20} + E_{27, 27} + E_{30, 30} - 
 E_{31, 31} + E_{32, 32} \nonumber\\ &+ E_{33, 33} - E_{34, 34} + E_{35, 35} - 
 E_{36, 36} - E_{37, 37} + E_{38, 38} - E_{40, 40} - E_{43, 43} + 
 E_{50, 50} + E_{54, 54} - E_{56, 56} + E_{57, 57} \nonumber\\ &+ E_{59, 59} - 
 E_{60, 60} + E_{61, 61} - E_{63, 63} + E_{64, 64} - E_{65, 65} + 
 E_{66, 66} - E_{67, 67} + E_{68, 68} - E_{70, 70} - E_{72, 72} + 
 E_{73, 73} \nonumber\\ &- E_{75, 75} - E_{80, 80} + E_{85, 85} - E_{92, 92} + 
 E_{93, 93} + E_{94, 94} + E_{98, 98} - E_{99, 99} - E_{100, 100} + 
 E_{101, 101} + E_{102, 102} \nonumber\\ &+ E_{104, 104} - E_{105, 105} - 
 E_{107, 107} - E_{108, 108} + E_{109, 109} + E_{110, 110} - 
 E_{111, 111} - E_{116, 116} + 2 E_{117, 117} - E_{118, 118}\nonumber\\ & + 
 E_{131, 131} - 2 E_{132, 132} + E_{133, 133} + E_{138, 138} - 
 E_{139, 139} - E_{140, 140} + E_{141, 141} + E_{142, 142} + 
 E_{144, 144} - E_{145, 145} \nonumber\\ &- E_{147, 147} - E_{148, 148} + 
 E_{149, 149} + E_{150, 150} - E_{151, 151} - E_{155, 155} - 
 E_{156, 156} + E_{157, 157} - E_{164, 164} + E_{169, 169}\nonumber\\ & + 
 E_{174, 174} - E_{176, 176} + E_{177, 177} + E_{179, 179} - 
 E_{181, 181} + E_{182, 182} - E_{183, 183} + E_{184, 184} - 
 E_{185, 185} + E_{186, 186}\nonumber\\ & - E_{188, 188} + E_{189, 189} - 
 E_{190, 190} - E_{192, 192} + E_{193, 193} - E_{195, 195} - 
 E_{199, 199} + E_{206, 206} + E_{209, 209} - E_{211, 211} \nonumber\\ &+ 
 E_{212, 212} + E_{213, 213} - E_{214, 214} + E_{215, 215} - 
 E_{216, 216} - E_{217, 217} + E_{218, 218} - E_{219, 219} - 
 E_{222, 222} + E_{229, 229}\nonumber\\ & - E_{232, 232} + E_{233, 233} - 
 E_{235, 235} + E_{236, 236} - E_{238, 238} + E_{244, 244} - 
 E_{245, 245},
\end{align}
\begin{align} 
 H_5&=E_{5, 5} - E_{6, 6} + E_{10, 10} - E_{11, 11} + E_{12, 12} - 
 E_{14, 14} + E_{20, 20} + E_{23, 23} - E_{24, 24} + E_{26, 26} - 
 E_{27, 27} + E_{29, 29} \nonumber\\ &- E_{30, 30} - E_{33, 33} + E_{36, 36} + 
 E_{40, 40} - E_{41, 41} + E_{43, 43} + E_{44, 44} - E_{45, 45} + 
 E_{48, 48} - E_{49, 49} - E_{50, 50} + E_{53, 53}\nonumber\\ & - E_{54, 54} - 
 E_{59, 59} + E_{63, 63} + E_{67, 67} - E_{69, 69} + E_{70, 70} + 
 E_{72, 72} - E_{74, 74} + E_{75, 75} - E_{76, 76} + E_{78, 78} - 
 E_{79, 79} \nonumber\\ &+ E_{80, 80} - E_{81, 81} - E_{85, 85} + E_{86, 86} - 
 E_{87, 87} + E_{91, 91} - E_{93, 93} + E_{97, 97} - E_{98, 98} + 
 E_{100, 100} + E_{103, 103}\nonumber\\ & - E_{104, 104} - E_{106, 106} + 
 E_{107, 107} + E_{108, 108} + E_{109, 109} - E_{110, 110} - 
 E_{113, 113} - E_{115, 115} + 2 E_{116, 116} \nonumber\\ &- E_{117, 117} + 
 E_{132, 132} - 2 E_{133, 133} + E_{134, 134} + E_{136, 136} + 
 E_{139, 139} - E_{140, 140} - E_{141, 141} - E_{142, 142} \nonumber\\ &+ 
 E_{143, 143} + E_{145, 145} - E_{146, 146} - E_{149, 149} + 
 E_{151, 151} - E_{152, 152} + E_{156, 156} - E_{158, 158} + 
 E_{162, 162} \nonumber\\ &- E_{163, 163} + E_{164, 164} + E_{168, 168} - 
 E_{169, 169} + E_{170, 170} - E_{171, 171} + E_{173, 173} - 
 E_{174, 174} + E_{175, 175} - E_{177, 177}\nonumber\\ &- E_{179, 179} + 
 E_{180, 180} - E_{182, 182} - E_{186, 186} + E_{190, 190} + 
 E_{195, 195} - E_{196, 196} + E_{199, 199} + E_{200, 200} - 
 E_{201, 201} \nonumber\\ &+ E_{204, 204} - E_{205, 205} - E_{206, 206} + 
 E_{208, 208} - E_{209, 209} - E_{213, 213} + E_{216, 216} + 
 E_{219, 219} - E_{220, 220} + E_{222, 222} \nonumber\\ &- E_{223, 223} + 
 E_{225, 225} - E_{226, 226} - E_{229, 229} + E_{235, 235} - 
 E_{237, 237} + E_{238, 238} - E_{239, 239} + E_{243, 243} - 
 E_{244, 244},
\end{align}
{\allowdisplaybreaks
\begin{align} 
 H_6&=E_{6, 6} - E_{7, 7} + E_{8, 8} - E_{10, 10} + E_{14, 14} + 
 E_{16, 16} - E_{17, 17} + E_{19, 19} - E_{20, 20} + E_{22, 22} - 
 E_{23, 23} + E_{25, 25} \nonumber\\ &- E_{26, 26} - E_{29, 29} + E_{41, 41} + 
 E_{45, 45} - E_{46, 46} + E_{49, 49} + E_{50, 50} - E_{51, 51} + 
 E_{54, 54} - E_{55, 55} + E_{56, 56} - E_{57, 57} \nonumber\\ &+ E_{59, 59} + 
 E_{60, 60} - E_{61, 61} - E_{63, 63} + E_{65, 65} - E_{66, 66} - 
 E_{67, 67} + E_{71, 71} - E_{72, 72} + E_{76, 76} - E_{78, 78} + 
 E_{81, 81} \nonumber\\ &- E_{83, 83} + E_{84, 84} + E_{87, 87} - E_{89, 89} + 
 E_{90, 90} - E_{91, 91} + E_{93, 93} - E_{95, 95} + E_{96, 96} - 
 E_{97, 97} + E_{99, 99} - E_{101, 101} \nonumber\\ &+ E_{102, 102} - 
 E_{103, 103} + E_{106, 106} - E_{107, 107} + E_{108, 108} - 
 E_{109, 109} - E_{114, 114} + 2 E_{115, 115} - E_{116, 116} + 
 E_{133, 133} \nonumber\\ &- 2 E_{134, 134} + E_{135, 135} + E_{140, 140} - 
 E_{141, 141} + E_{142, 142} - E_{143, 143} + E_{146, 146} - 
 E_{147, 147} + E_{148, 148} - E_{150, 150} \nonumber\\ &+ E_{152, 152} - 
 E_{153, 153} + E_{154, 154} - E_{156, 156} + E_{158, 158} - 
 E_{159, 159} + E_{160, 160} - E_{162, 162} - E_{165, 165} + 
 E_{166, 166} \nonumber\\ &- E_{168, 168} + E_{171, 171} - E_{173, 173} + 
 E_{177, 177} - E_{178, 178} + E_{182, 182} + E_{183, 183} - 
 E_{184, 184} + E_{186, 186} + E_{188, 188} \nonumber\\ &- E_{189, 189} - 
 E_{190, 190} + E_{192, 192} - E_{193, 193} + E_{194, 194} - 
 E_{195, 195} + E_{198, 198} - E_{199, 199} - E_{200, 200} + 
 E_{203, 203} \nonumber\\ &- E_{204, 204} - E_{208, 208} + E_{220, 220} + 
 E_{223, 223} - E_{224, 224} + E_{226, 226} - E_{227, 227} + 
 E_{229, 229} - E_{230, 230} + E_{232, 232} \nonumber\\ &- E_{233, 233} - 
 E_{235, 235} + E_{239, 239} - E_{241, 241} + E_{242, 242} - 
 E_{243, 243},
\end{align}}
\begin{align} 
 H_7&=E_{7, 7} - E_{9, 9} + E_{10, 10} + E_{11, 11} - E_{12, 12} + 
 E_{13, 13} - E_{14, 14} + E_{15, 15} - E_{16, 16} + E_{18, 18} - 
 E_{19, 19} + E_{21, 21} \nonumber\\ &- E_{22, 22} - E_{25, 25} + E_{46, 46} + 
 E_{51, 51} - E_{52, 52} + E_{55, 55} + E_{57, 57} - E_{58, 58} + 
 E_{61, 61} - E_{62, 62} + E_{63, 63} - E_{64, 64} \nonumber\\ &+ E_{66, 66} + 
 E_{67, 67} - E_{68, 68} + E_{69, 69} - E_{70, 70} + E_{72, 72} - 
 E_{73, 73} + E_{74, 74} - E_{75, 75} - E_{76, 76} + E_{77, 77} + 
 E_{78, 78}\nonumber\\ & + E_{79, 79} - E_{80, 80} - E_{81, 81} + E_{82, 82} - 
 E_{84, 84} + E_{85, 85} - E_{86, 86} - E_{87, 87} + E_{88, 88} - 
 E_{90, 90} + E_{92, 92} - E_{93, 93}\nonumber\\ & + E_{94, 94} - E_{96, 96} - 
 E_{99, 99} + E_{100, 100} - E_{102, 102} + E_{106, 106} - 
 E_{108, 108} + 2 E_{114, 114} - E_{115, 115} + E_{134, 134}\nonumber\\ & - 
 2 E_{135, 135} + E_{141, 141} - E_{143, 143} + E_{147, 147} - 
 E_{149, 149} + E_{150, 150} + E_{153, 153} - E_{155, 155} + 
 E_{156, 156} - E_{157, 157} \nonumber\\ &+ E_{159, 159} - E_{161, 161} + 
 E_{162, 162} + E_{163, 163} - E_{164, 164} + E_{165, 165} - 
 E_{167, 167} + E_{168, 168} + E_{169, 169} - E_{170, 170}\nonumber\\ & - 
 E_{171, 171} - E_{172, 172} + E_{173, 173} + E_{174, 174} - 
 E_{175, 175} + E_{176, 176} - E_{177, 177} + E_{179, 179} - 
 E_{180, 180} + E_{181, 181}\nonumber\\ & - E_{182, 182} - E_{183, 183} + 
 E_{185, 185} - E_{186, 186} + E_{187, 187} - E_{188, 188} + 
 E_{191, 191} - E_{192, 192} - E_{194, 194} + E_{197, 197} \nonumber\\ &- 
 E_{198, 198} - E_{203, 203} + E_{224, 224} + E_{227, 227} - 
 E_{228, 228} + E_{230, 230} - E_{231, 231} + E_{233, 233} - 
 E_{234, 234} + E_{235, 235} \nonumber\\ &- E_{236, 236} + E_{237, 237} - 
 E_{238, 238} - E_{239, 239} + E_{240, 240} - E_{242, 242},
\end{align}
{\allowdisplaybreaks
\begin{align}
 H_8&=E_{6, 6} + E_{7, 7} - E_{8, 8} + E_{9, 9} - E_{10, 10} - E_{12, 12} + 
 E_{24, 24} + E_{27, 27} - E_{28, 28} + E_{30, 30} + E_{31, 31} - 
 E_{32, 32} + E_{33, 33} \nonumber\\ &+ E_{34, 34} - E_{35, 35} - E_{36, 36} + 
 E_{37, 37} - E_{38, 38} + E_{39, 39} - E_{40, 40} + E_{42, 42} - 
 E_{43, 43} - E_{44, 44} + E_{47, 47} - E_{48, 48}\nonumber\\ & - E_{53, 53} + 
 E_{69, 69} + E_{74, 74} + E_{76, 76} - E_{77, 77} + E_{79, 79} + 
 E_{81, 81} - E_{82, 82} + E_{83, 83} - E_{84, 84} + E_{85, 85} + 
 E_{87, 87} \nonumber\\ &- E_{88, 88} + E_{89, 89} - E_{90, 90} - E_{91, 91} + 
 E_{92, 92} + E_{93, 93} - E_{94, 94} + E_{95, 95} - E_{96, 96} - 
 E_{97, 97} + E_{99, 99} - E_{100, 100} \nonumber\\ &+ E_{101, 101} - 
 E_{102, 102} - E_{103, 103} + E_{107, 107} - E_{108, 108} - 
 E_{109, 109} + 2 E_{113, 113} - E_{116, 116} + E_{133, 133} - 
 2 E_{136, 136} \nonumber\\ &+ E_{140, 140} + E_{141, 141} - E_{142, 142} + 
 E_{146, 146} + E_{147, 147} - E_{148, 148} + E_{149, 149} - 
 E_{150, 150} + E_{152, 152} + E_{153, 153} \nonumber\\ &- E_{154, 154} + 
 E_{155, 155} - E_{156, 156} - E_{157, 157} + E_{158, 158} + 
 E_{159, 159} - E_{160, 160} + E_{161, 161} - E_{162, 162} - 
 E_{164, 164}\nonumber\\ & + E_{165, 165} - E_{166, 166} + E_{167, 167} - 
 E_{168, 168} - E_{170, 170} + E_{172, 172} - E_{173, 173} - 
 E_{175, 175} - E_{180, 180} + E_{196, 196} \nonumber\\ &+ E_{201, 201} - 
 E_{202, 202} + E_{205, 205} + E_{206, 206} - E_{207, 207} + 
 E_{209, 209} - E_{210, 210} + E_{211, 211} - E_{212, 212} + 
 E_{213, 213} \nonumber\\ &+ E_{214, 214} - E_{215, 215} - E_{216, 216} + 
 E_{217, 217} - E_{218, 218} - E_{219, 219} + E_{221, 221} - 
 E_{222, 222} - E_{225, 225} + E_{237, 237} \nonumber\\ &+ E_{239, 239} - 
 E_{240, 240} + E_{241, 241} - E_{242, 242} - E_{243, 243}.
\end{align}
}}

\subsection{\texorpdfstring{$B_r$}{Br}}

In the $(2r+1)$-dimensional representation of $B_r$, the Cartan generators are
\begin{align}
    H_i&=E_{i,i}-E_{i+1,i+1}+E_{2r+1-i,2r+1-i}-E_{2r+2-i,2r+2-i}, \quad i=1,\cdots, r-1,\nonumber\\
    H_r&=E_{r,r}-E_{r+2,r+2}.
\end{align}

\subsection{\texorpdfstring{$F_4$}{F4}}

The Cartan generators in the 26-dimensional representation are
{\allowdisplaybreaks\small
\begin{align}
    H_1&=E_{4,4}+E_{5,5}-E_{6,6}+E_{7,7}-E_{8,8}-E_{9,9}+E_{17,17}-E_{19,19}+E_{20,20}-E_{21,21}+E_{22,22}-E_{23,23},
    \nonumber\\
    H_2&=E_{3,3}-E_{4,4}+E_{8,8}+E_{9,9}-E_{10,10}-E_{11,11}+E_{15,15}-E_{17,17}+E_{18,18}-E_{20,20}+E_{23,23}-E_{24,24},
    \nonumber\\
    2H_3&=E_{2,2}-E_{3,3}+E_{4,4}-E_{5,5}+E_{6,6}-E_{8,8}+E_{10,10}+2E_{11,11}-E_{12,12}-2E_{15,15}+E_{16,16}
    \nonumber\\
    &\quad -E_{18,18}+E_{20,20}+E_{21,21}-E_{22,22}-E_{23,23}+E_{24,24}-E_{25,25},
    \nonumber\\
    2H_4&=E_{1,1}-E_{2,2}+E_{5,5}-E_{7,7}+E_{8,8}-E_{9,9}+E_{10,10}-E_{11,11}+2E_{12,12}+E_{15,15}-2E_{16,16}
    \nonumber\\
    &\quad +E_{17,17}-E_{18,18}+E_{19,19}-E_{20,20}-E_{21,21}+E_{25,25}-E_{26,26}.
\end{align}
}
We change the ordering of the simple roots from \cite{Ito:2020htm}.

\section{Recursion Relations in \texorpdfstring{$\mathcal{W}D_r$}{WDr} Algebra}
\label{app:WD-recursion}

In this appendix, we present some recursion relations for the generators in the $\W D_r$ algebra, which provide the $\W$-currents up to spin 8 for general rank $r$.
In the following formulas, we define $b[j]:=1-2j(2j+1)Q^2$.

\begin{equation}
\label{eq:wdrw2decomposition}
    W_{2}^{(r)} =
    W^{(r-1)}_2 + \frac{1}{2}(\tp_r \tp_r) - (r-1)Q\partial \tp_r,
\end{equation}
\begin{align}
    W_4^{(r)}
    &=
    W_4^{(r-1)} 
    + (\tp_r(\tp_r W^{(r-1)}_2))
    - 2(r-2)Q (\partial \tp_r W_2^{(r-1)})
    - Q (\tp_r \partial W_2^{(r-1)})
    \nonumber\\
    &\quad
    + \frac{1}{4}(1-Q^2(2r-4)(2r-5)) \partial^2 W_2^{(r-1)}
    - \frac{1}{4}
    b[r-2] \partial^2 W_2^{(r)}
    \nonumber\\
    &\quad + \frac{1}{12}b[r-2](3(\tp_r\partial^2 \tp_r)-Q(2r-2)\partial^3 \tp_r),
\end{align}
{\small
\begin{align}
    W_6^{(r)}
    &=
    W_6^{(r-1)}
    + (\tp_r(\tp_r W_4^{(r-1)}))
    - Q(2r-6)(\partial \tp_r W_4^{(r-1)})
    - Q(\tp_r \partial W_4^{(r-1)})
    \nonumber\\
    &\quad
    + \frac{1}{4}(1-Q^2(2r-6)(2r-7))\partial^2 W_4^{(r-1)}
    + \frac{b[r-3]}{6}
    \qty(
    3(\tp_r(\partial^2\tp_r W_2^{(r-1)}))
    - Q(2r-4)(\partial^3\tp_r W_2^{(r-1)})
    )
    \nonumber\\
    &\quad
    + \frac{b[r-3]}{4}\Bigl(2 (\tp_r(\partial \tp_r \partial W_2^{(r-1)}))
    - Q(2r-3)(\partial^2 \tp_r\partial W_2^{(r-1)})
    \Bigr)
    \nonumber\\
    &\quad
    + \frac{b[r-3]}{4}
    \qty(
    (\tp_r(\tp_r \partial^2 W_2^{(r-1)})) - Q(2r-4)(\partial\tp_r\partial^2 W_2^{(r-1)} )
    )
    - \frac{Q}{4} b[r-3] (\tp_r \partial^3 W_2^{(r-1)})
    \nonumber\\
    &\quad
    + (1-Q^2 (2r-4)(2r-7))\frac{b[r-3]}{48} \partial^4 W_2^{(r-1)}
    + \frac{b[r-2]b[r-3]}{240}
    \qty(
    5 (\tp_r \partial^4 \tp_r) - Q(2r-2) \partial^5 \tp_r)
    \nonumber\\
    &\quad
    - \frac{b[r-3]}{4} \partial^2 W_4^{(r)}-\frac{b[r-2]b[r-3]}{48}\partial^4 W_2^{(r)},
\end{align}
{\allowdisplaybreaks
\begin{align}
    W_8^{(r)}&=W_8^{(r-1)}
    \nonumber\\
    &\quad +(\tp_r(\tp_r W_6^{(r-1)})) -Q(2r-8) (\partial\tp_r W_6^{(r-1)})
    -Q(\tp_r \partial W_6^{(r-1)})
    +\frac{1}{4}(1-Q^2(2r-9)(2r-8))\partial^2 W_6^{(r-1)}
    \nonumber\\
    &\quad +\frac{b[r-4]}{6}
    \Bigl(
    3 (\tp_r(\partial^2\tp_r W_4^{(r-1)}))
    -Q(2r-6)(\partial^3\tp_r W_4^{(r-1)})
    \Bigr)
    \nonumber\\
    &\quad +\frac{b[r-4]}{4}
    \Bigl(
    2(\tp_r(\partial \tp_r \partial W_4^{(r-1)}))
    -Q(2r-5) (\partial^2\tp_r\partial W_4^{(r-1)})
    \Bigr)
    \nonumber\\
    &\quad +\frac{b[r-4]}{4}
    \Bigl(
    (\tp_r(\tp_r \partial^2 W_4^{(r-1)}))
    -Q(2r-6) (\partial\tp_r\partial^2 W_4^{(r-1)})
    \Bigr)
    \nonumber\\
    &\quad -\frac{Q b[r-4]}{4} (\tp_r \partial^3 W_4^{(r-1)})
    +\frac{b[r-4]}{48}(1-Q^2(2r-6)(2r-9))\partial^4 W_4^{(r-1)}
    \nonumber\\
    &\quad +\frac{b[r-3]b[r-4]}{120}
    \Bigl(
    5 (\tp_r(\partial^4 \tp_r W_2^{(r-1)}))-Q(2r-4)(\partial^5\tp_r W_2^{(r-1)})
    \Bigr)
    \nonumber\\
    &\quad +\frac{b[r-3]b[r-4]}{48}
    \Bigl(
    4 (\tp_r(\partial^3 \tp_r\partial W_2^{(r-1)})) - Q(2r-3)(\partial^4\tp_r \partial W_2^{(r-1)})
    \Bigr)
    \nonumber\\
    &\quad +\frac{b[r-3]b[r-4]}{24}
    \Bigl(
    3 (\tp_r(\partial^2 \tp_r\partial^2 W_2^{(r-1)}))-Q(2r-4)(\partial^3\tp_r \partial^2 W_2^{(r-1)})
    \Bigr)
    \nonumber\\
    &\quad +\frac{b[r-3]b[r-4]}{12}
    \Bigl(
     (\tp_r(\partial \tp_r\partial^3 W_2^{(r-1)}))-Q(r-2)(\partial^2\tp_r \partial^3 W_2^{(r-1)})
    \Bigr)
    \nonumber\\
    &\quad +\frac{b[r-3]b[r-4]}{48}
    \Bigl(
     (\tp_r( \tp_r\partial^4 W_2^{(r-1)}))-Q(2r-4)(\partial\tp_r \partial^4 W_2^{(r-1)})
    \Bigr)
    \nonumber\\
    &\quad -\frac{Q b[r-3]b[r-4]}{48} (\tp_r \partial^5 W_2^{(r-1)})
    +\frac{b[r-3]b[r-4]}{1440}(1-Q^2(2r-4)(2r-9) )\partial^6 W_2^{(r-1)}
    \nonumber\\
    &\quad -\frac{b[r-4]}{4}\partial^2 W_6^{(r)}-\frac{b[r-3]b[r-4]}{48}\partial^4 W_4^{(r)}-\frac{b[r-2]b[r-3]b[r-4]}{1440}\partial^6 W_2^{(r)}
    \nonumber\\
    &\quad +\frac{b[r-2]b[r-3]b[r-4]}{10080}
    \Bigl(
    7 (\tp_r \partial^6 \tp_r)-Q(2r-2)\partial^7\tp_r
    \Bigr).
\end{align}
}
}
The initial conditions with $r=1$ is given by 
{\allowdisplaybreaks\small
\begin{align}
    W_2^{(1)} &= \frac{1}{2}(\tp_1\tp_1),\\
    W_4^{(1)} &= \frac{1}{4}(1-2Q^2) \qty((\partial^2\tp_1\tp_1)-\frac{1}{2}\partial^2(\tp_1\tp_1)),
     \\
    W_6^{(1)}
    &=
    - \frac{1}{4}b[-2]\partial^2 W^{(1)}_4 + \frac{1}{48}b[-1]b[-2] 
    (\partial^4 \tp_1\tp_1)
    - \frac{1}{48}b[-1]b[-2] \partial^4 W_2^{(1)}, \\
    W_8^{(1)}
    &=
    - \frac{1}{4} b[-3]\partial^2 W_6^{(1)}
    - \frac{1}{48}b[-2]b[-3]\partial^4 W_4^{(1)} -\frac{1}{1440}b[-1]b[-2]b[-3]\partial^6 W_2^{(1)}
    + \frac{1}{1440}b[-1]b[-2]b[-3](\partial^6 \tp_1 \tp_1).
\end{align}
}

\section{\texorpdfstring{$\mathcal{W}$}{W}-charge for \texorpdfstring{$W_{12}$}{W12} in \texorpdfstring{$\mathcal{W} E_7$}{WE7} Algebra}\label{app:WE7Charge}

The $\W$-charge for the spin-$12$ generator $W_{12}$ in \eqref{eq:WE7currents} is expressed in terms of the Casimir invariants $C_k$ as follows:

{\small\allowdisplaybreaks
\begin{align}
    \Delta_{12} &= \frac{2C_{12}}{10125}(25 - 414Q^2) - \frac{C_{10}C_2}{1020600}(1009 - 5868Q^2) + \frac{C_{10}}{94500}(108400 - 12701465Q^2 - 80398044Q^4)
    \nonumber\\
    &\quad 
    + \frac{C_8C_2^2}{81648000}(5125 + 35586Q^2) - \frac{C_8C_2}{1134000}(206100 - 19870991Q^2 - 162953073Q^4)
    \nonumber\\
    &\quad + \frac{C_8}{36000}(2393000 - 438088080Q^2 + 3443508877Q^4 + 56248726446Q^6) - \frac{C_6^2}{291600}(103 + 1908Q^2)
    \nonumber\\
    &\quad
    + \frac{C_6C_2^3}{3359232000}(3679 + 55206Q^2) + \frac{C_6C_2^2}{62208000}(745528 - 22157923Q^2 - 1132037822Q^4)
    \nonumber\\
    &\quad - \frac{C_6C_2}{31104000}(191426796 - 51246586128Q^2 + 970799935193Q^4 + 98558165796282Q^6)
    \nonumber\\
    &\quad + \frac{C_6}{1728000}(1242741312 - 284712425444Q^2 + 27544402888824Q^4 - 805585336841377Q^6
    \nonumber\\
    &\qquad\qquad\qquad - 116944368153195690Q^8)
    \nonumber\\
    &\quad - \frac{C_2^6}{3869835264000}(13757 + 265122Q^2) - \frac{C_2^5}{107495424000}(1402035 + 45135254Q^2 - 3582170400Q^4)
    \nonumber\\
    &\quad + \frac{C_2^4}{35831808000}(152314404 - 75412965102Q^2 + 2397313642535Q^4 + 271850187397518Q^6)
    \nonumber\\
    &\quad - \frac{C_2^3}{995328000}(1656267848 - 368442769692Q^2 + 33609513390706Q^4 - 1139286281354991Q^6
    \nonumber\\
    &\qquad\qquad\qquad\quad - 96900821845766142Q^8)
    \nonumber\\
    &\quad - \frac{C_2^2}{331776000}(2788289760 - 2259127069968Q^2 + 705931690267176Q^4 - 89693625388292364Q^6
    \nonumber\\
    &\qquad\qquad\qquad\quad - 379584226105254463Q^8 + 564524747181454240650Q^{10})
    \nonumber\\
    &\quad - \frac{C_2}{27648000}(10655205280 - 31431794042016Q^2 + 10940750146178520Q^4 - 729503123599951240Q^6 
    \nonumber\\
    &\qquad\qquad\qquad\quad - 71942133358433209539Q^8 + 2777893837350370036625Q^{10} + 34212076717820285104326Q^{12})
    \nonumber\\
    &\quad + \frac{19Q^2}{3072000}(149172873920 - 432257423288544Q^2 + 146923401368380080Q^4 - 8228913425582484156Q^6
    \nonumber\\
    &\qquad\qquad\qquad - 1262881954272457490466Q^8 + 37594076868389667176977Q^{10}
    \nonumber\\
    &\qquad\qquad\qquad + 2177963721238515196592610Q^{12}).
\end{align}}

\section{\texorpdfstring{$\mathcal{W}$}{W}-charges for \texorpdfstring{$W_8$}{W8} and \texorpdfstring{$W_{12}$}{W12} in \texorpdfstring{$\mathcal{W}F_4$}{WF4} Algebra}\label{app:WF4Charge}

The $\W$-charges for the spin-8 and spin-12 generators $W_8, W_{12}$ in \eqref{eq:WF4W8W12} are expressed as follows:
{\allowdisplaybreaks
\small
\begin{align}
    \Delta_8
    &= - \frac{125\left(9b^2-17\right)\left(17b^2-18\right)}{72b^2}C_8 + \frac{5\left(12383b^4-36515b^2+25160\right)}{864b^2}C_6C_2
    \nonumber\\
    &\quad -\frac{966499b^4-2877125b^2+2028490}{2488320b^2}C_2^4
    \nonumber\\
    &\quad -\frac{5\left(1104831b^8-2899229b^6-393063b^4+6048655b^2-4017780\right)}{216b^4}C_6
    \nonumber\\
    &\quad +\frac{826337433b^8-3871585060b^6+6530404880b^4-4552343125b^2+1016205600}{1555200b^4}C_2^3
    \nonumber\\
    &\quad -\frac{1}{259200b^6}\left(76542215247b^{12}-742310074855b^{10}+2983521002769b^8-6342309218155b^6\right.
    \nonumber\\
    &\qquad\qquad\qquad\quad \left.+ 7511609153520b^4-4698567074200b^2+1213046558100\right)C_2^2
    \nonumber\\
    &\quad -\frac{1}{14400b^8}\left(-2009758474374b^{16}+22846225158765b^{14}-117994681013937b^{12}\right.
    \nonumber\\
    &\qquad\qquad\qquad \left.+358292271546140b^{10}-692990395642976b^8+867264498691700b^6\right.
    \nonumber\\
    &\qquad\qquad\qquad \left.-681766613438960b^4+306572026645500b^2-60227045023200\right)C_2
    \nonumber\\
    &\quad - \frac{1}{2400b^{10}}\left(48982658395086b^{20}-653751148890130b^{18}+4066232113256145b^{16}\right.
    \nonumber\\
    &\qquad\qquad\qquad \left.-15499494046350445b^{14}+39922035556094640b^{12}-72163767996309600b^{10}\right.
    \nonumber\\
    &\qquad\qquad\qquad \left.+92119690866294868b^8-81531394239173700b^6+47660617668387060b^4\right.
    \nonumber\\
    &\qquad\qquad\qquad \left.-16560003571723900 b^2+2590921525631400\right),
\end{align}
\begin{align}
    \nonumber
    \Delta_{12}&=-\frac{3125\left(9b^2-17\right) \left(17b^2-18\right)}{81b^2}C_{12}+\frac{125 \left(381033b^4-1123297b^2+765612\right)}{93312b^2}C_8C_2^2\\
    \nonumber
    &+\frac{125\left(120141b^4-354505b^2+250920\right)}{23328b^2}C_6^2-\frac{5\left(66921111b^4-198031165b^2+136611990\right)}{3359232b^2}C_6C_2^3\\
    \nonumber
    &+\frac{212072823b^4-630013625b^2+438092730}{537477120b^2}C_2^6\\
    \nonumber
    &-\frac{25\left(11744073b^8-22751901b^6-33669119b^4+98957105b^2-56420280\right)}{1296b^4}C_8C_2\\
    \nonumber
    &+\frac{5\left(3545224320b^8-17063459871b^6+30071418557b^4-22566800370b^2+5768693250\right)}{93312b^4}C_6C_2^2\\
    \nonumber
    &-\frac{182153419077b^8-945694878616b^6+1814747656412b^4-1507169334355b^2+445817497470}{134369280b^4}C_2^5\\
    \nonumber
    &+\frac{25}{432b^6}\left(12309561144b^{12}-69599436615b^{10}+123596653785b^8-9369982420b^6\right.\\
    \nonumber
    &\left.\qquad\qquad -215096722694b^4+240365465970b^2-82308926520\right)C_8\\
    \nonumber
    &-\frac{1}{15552b^6}\left(6739646999238b^{12}-57134481847716b^{10}+205300096742921b^8-398429835676035b^6\right.\\
    \nonumber
    &\left.\qquad\qquad\quad +438676715956820b^4-259048394460350b^2+64019033693100\right)C_6C_2\\
    \nonumber
    &+\frac{1}{559872000b^6}\left(1451193192177348b^{12}-12740818639618050b^{10}+47486795189805395b^8\right.\\
    \nonumber
    &\left.\qquad\qquad\qquad\quad -95425540009724675b^6+108440994142067725b^4-65876359369354375b^2\right.\\
    \nonumber
    &\left.\qquad\qquad\qquad\quad +16702443753930000\right)C_2^4\\
    \nonumber
    &+\frac{1}{2592b^8}\left(1457873225405982b^{16}-13249527571839222b^{14}+56040116399971053b^{12}\right.\\
    \nonumber
    &\left.\qquad\qquad\quad -147352687983797599b^{10}+264492587363786638b^8-326233826710848550b^6\right.\\
    \nonumber
    &\left.\qquad\qquad\quad +262919027998020880b^4-123624216236645400b^2+25558383124762200\right)C_6\\
    \nonumber
    &-\frac{1}{93312000b^8}\left(377128793166027426b^{16}-2985235533987791189b^{14}+10093085001890898637 b^{12}\right.\\
    \nonumber
    &\left.\qquad\qquad\qquad -19223175745308970990b^{10}+23199737103589507360b^8-19274803044828893750b^6\right.\\
    \nonumber
    &\left.\qquad\qquad\qquad +11653433683548893700b^4-4854581852192325000b^2+1015041145571640000\right)C_2^3\\
    \nonumber
    &-\frac{1}{15552000b^{10}}\left(22602987976163301936b^{20}-324640207773457732350b^{18}\right.\\
    \nonumber
    &\left.\qquad\quad +2163981475420767801999b^{16}-8873197544120415678145b^{14}+24748825070794492280944b^{12}\right.\\
    \nonumber
    &\left.\qquad\quad -48700946247887415007330b^{10}+67808032956666717214840b^8-65361337557325544163700b^6\right.\\
    \nonumber
    &\left.\qquad\quad +41449474319908827730600b^4-15541629193651019484000b^2+2608977174358536270000\right)C_2^2\\
    \nonumber
    &+\frac{1}{864000b^{12}}\left(255238062149830458366b^{24}-5393138896610153753529b^{22}\right.\\
    \nonumber
    &\left.+48406377614945305794788b^{20}-256508003671421787672123b^{18}+916319507820616158972446b^{16}\right.\\
    \nonumber
    &\left.-2353801931676883338903616b^{14}+4477295493904906530261588b^{12}\right.\\
    \nonumber
    &\left.-6347635868627518692058120b^{10}+6631529738292495575204000b^8-4955895684371497253840600b^6\right.\\
    \nonumber
    &\left.+2504280347705584957931600b^4-765776414262780291398000b^2+106925152163996615652000\right)C_2\\
    \nonumber
    &+\frac{1}{288000b^{14}}\left(552873688500109442940b^{28}+77921186936009879593593b^{26}-1336146068084008059403581b^{24}\right.\\
    \nonumber
    &\left.+10357761482131938309925874b^{22}-49778547228581002295861684b^{20}+167007910447154233950537308b^{18}\right.\\
    \nonumber
    &\left.-414456508582166835004291424b^{16}+782696994455958340564081624b^{14}-1136040029654831561508323840b^{12}\right.\\
    \nonumber
    &\left.+1262191802796240238041851120b^{10}-1055313731179283240625892960b^8+642518925057136842990252800b^6\right.\\
    &\left.-268718344092602870261804000b^4+68981231588658792125820000b^2-8189800239140812785024000\right).
\end{align}
}

\bibliographystyle{JHEP}
\bibliography{refs.bib}

\end{document}